\documentclass[aps,11pt,superscriptaddress,footinbib]{revtex4-1}
\usepackage{amssymb}
\usepackage{amsmath}
\newcommand\new[1]{#1}
\renewcommand\min{\!-\!}

\begin{document}
\title{Gravitational dynamics for all tensorial spacetimes\\
carrying predictive, interpretable and quantizable matter}
\author{Kristina Giesel}
\address{Universit\"at Erlangen, Institut f\"ur Theoretische Physik III, Lehrstuhl f\"ur Quantengravitation, Staudtstr. 7, 91058 Erlangen, Germany}
\author{Frederic P. Schuller}\email{Corresponding author: fps@aei.mpg.de}
\address{Albert--Einstein--Institut\\Max--Planck--Institut f\"ur Gravitationsphysik\\Am M\"uhlenberg 1, 14476 Potsdam, Germany}
\author{Christof Witte}
\address{Albert--Einstein--Institut\\Max--Planck--Institut f\"ur Gravitationsphysik\\Am M\"uhlenberg 1, 14476 Potsdam, Germany}
\author{Mattias N. R. Wohlfarth}
\address{Zentrum f\"ur Mathematische Physik und II. Institut f\"ur Theoretische Physik\\
Universit\"at Hamburg, Luruper Chaussee 149, 22761 Hamburg, Germany}

\begin{abstract}
Only a severely restricted class of tensor fields can provide
classical spacetime geometries, namely those that can carry matter field equations that are predictive, interpretable and quantizable. These three conditions on matter translate into three corresponding algebraic conditions on the underlying tensorial geometry, namely to be hyperbolic, time-orientable and energy-distinguishing. Lorentzian metrics, on which general relativity and the standard model of particle physics are built, present just the simplest tensorial spacetime geometry satisfying these conditions. 
The problem of finding gravitational dynamics---for the general tensorial spacetime geometries satisfying the above minimum requirements---is reformulated in this paper as a system of linear partial differential equations, in the sense that their solutions yield the actions governing the corresponding spacetime geometry. Thus the search for modified gravitational dynamics is reduced to a clear mathematical task. 
\end{abstract}
\date\today
\maketitle
\tableofcontents
\newpage
\section{Introduction}
  Over the past two decades, remarkable gaps in our understanding of matter have been revealed---astrophysical observations \cite{WMAP} by now overwhelmingly indicate that only about 4\% of the matter and energy in the universe can be of standard model origin. Indeed, one cannot attribute the remaining 21\% of dark matter or 75\% of dark energy to standard model matter or its vacuum energy. For on the one hand, does the bullet cluster \cite{bullet} show the existence of only gravitationally interacting dark matter. On the other hand does the calculation of dark energy as the vacuum energy of standard model fields yield a result that is infamously off the observed value by 120 order of magnitude \cite{cosmological}, which is jokingly referred to as the worst prediction of elementary particle physics. \new{It is consensus that these observations thus point at something fundamental we do not understand about matter or gravity.

Going deeper than just postulating modified Lagrangians for either matter or gravity, however, one quickly realizes that modifying matter and gravity dynamics independently quickly becomes inconsistent. This is because they both build on---and more importantly, are both tightly constrained by---the common underlying spacetime geometry.

For the Einstein equations, on the one hand, present the unique dynamics with a well-posed initial value problem one can give to a Lorentzian manifold \cite{Kuchar}. Thus modifying gravitational dynamics necessarily comes at the cost of deviating from Lorentzian geometry as the spacetime structure. This is illustrated for instance by Brans-Dicke gravity and its avatars as well as $f(R)$ gravity theories, which all feature at least an additional scalar gravitational degree of freedom. But then one needs to couple matter fields to the corresponding modified spacetime geometry, and one sees that modifying gravitational dynamics compels one to think about modified matter dynamics.   

Vice versa, even minimal deviations from the standard model of particle physics on the other hand quickly produce matter dynamics whose causality does not coincide with the causality defined by the underlying Lorentzian manifold. Famously, this surprisingly already happens  for some fully covariant Lagrangians \cite{VeloZwanziger}. If one wishes to consider such matter dynamics on a Lorentzian manifold, let alone any even slightly more exotic matter, one requires also modified gravity dynamics to provide backgrounds that make the matter equations causal. Thus unless one entertains the claim that all matter that could exist in Nature must be made precisely in the image of Maxwell theory---which in the face of 96\% of all matter and energy in the universe being of entirely unknown origin appears an unnecessarily restrictive and arbitrary idea---one sees that modifying matter dynamics generically compels one to construct a corresponding modified gravity theory to be consistent.  

Therefore if one sets out to modify the otherwise time-tested theory of Einstein gravity coupled to standard model dynamics, the intimate link between consistent matter dynamics and gravitational field equations compels one to be particularly careful, and base any modification on a common underlying geometry that is so constrained as to make the combined theory work.
At the very least, it is to be physically required that the spacetime geometry render the entirety of observed matter field dynamics coupling to it predictive,  interpretable and quantizable \footnote{The precise technical meaning of predictivity, interpretability and quantizability is discussed in \cite{RRS} and reviewed, to the extent necessary for a full understanding of the present paper, in section \ref{sec_primer}}. The identification of all tensorial geometries complying with these minimum criteria and the determination of their gravitational dynamics is the purpose of the present paper.\enlargethispage{1cm}

Fortunately, these rather fundamental physical conditions translate into three simple algebraic conditions \cite{RRS} that an otherwise arbitrary tensor field must satisfy in order to provide a valid spacetime geometry:} it must be hyperbolic, time-orientable and energy-distinguishing, as we will explain in the first technical section below. Thus the spectrum of tensor fields that can serve as a spacetime structure in the presence of specific matter field dynamics \footnote{Having to invoke specific matter field equations in order to decide on what constitutes a spacetime geometry only seems odd at first sight. For in fact this is precisely the reasoning that led Einstein from Maxwell electrodynamics---having properties that at the time contradicted the conceived  model of space and time as much as superluminal neutrinos do now---ultimately to the Lorentzian geometry of spacetime.} is greatly restricted. So restricted, in fact, that all kinematical constructions familiar from the special case of Lorentzian metrics can be made, for precisely the same structural reasons, also for any such tensorial spacetime.

The deeper dynamical principles behind Einstein's field equations, which were revealed by Arnowitt, Deser and Misner \cite{ADM} a long time ago, are fortunately not swept away with a change from Lorentzian geometry to one of the alternative tensorial geometries described above. From the geometrodynamical point of view, gravitational dynamics is all about evolving the spatial geometry from one  suitable initial data surface to an infinitesimally neighbouring one, such that ultimately all spatial geometries recombine to an admissible spacetime geometry; indeed, applying this principle, Hojmann, Kuchar and Teitelboim \cite{HKT, Kuchar} derived the Einstein-Hilbert action with numerically undetermined gravitational and cosmological constants as the unique dynamics for the special case when the tensorial geometry is in fact a Lorentzian metric. But this geometrodynamic principle stands 
for any tensorial spacetime geometry \footnote{As long as one understands the Legendre duality map between co-vectors that are normal to admissible initial data surfaces and the vectors they correspond to, which is derived in \cite{RRS} and concisely reviewed in section \ref{sec_primer}}. And it is the purpose of the present article to show that the very same principle indeed determines the dynamics of any hyperbolic, time-orientable and energy-distinguishing tensorial spacetime geometry. But indeed only of such. 

The main result of this article, beyond its technical details, is thus the observation that the search for gravitational dynamics beyond general relativity can be reduced to solving a mathematical representation problem. This is achieved by invoking precisely the same geometrodynamic principles as followed by \cite{HKT}, but applied to any hyperbolic, time-orientable and energy-distinguishing tensorial spacetime geometry. As a consequence, gravitational dynamics (including and beyond \footnote{The history of modified gravity theories is a long and winding one, from the recently heavily studied $f(R)$ modified gravity actions for Lorentzian manifolds, over various actions for non-symmetric or scalar-tensor extensions of the gravitational field, dynamics for the affine connection rather than a metric, and finally to string-geometry inspired proposals of various sorts. All these proposals share the pleasant feature of being inspired in some way; compelling, however, let alone without alternative in the sense of the present paper, appear few of these classes of theories.} Einstein) need not be postulated, since they can be derived for any tensorial spacetime geometry, as we will show, by solving a
a system of homogeneous linear partial differential equations.   
Thus the question whether there is an alternative to general relativity as a classical gravity theory carries over in the formalism developed in the paper to a mathematical existence problem. Now if such solutions exist at all, the question of whether there are other dynamics for the same geometry translates into the mathematical question of the uniqueness of such a solution. And finally, the problem of constructing concrete alternative gravity dynamics amounts to nothing less, but also nothing more, than finding an actual solution of the linear system of partial differential equations. In the philosophy of this paper, studying modified theories of gravity therefore amounts to studying solutions to the said equations.

The organisation of this paper is as follows. In section \ref{sec_primer}, we start by giving a concise review of tensorial spacetime geometries, culminating in the insight of how normal co-vectors to initial data surfaces are mapped to normal vectors, which is the pivotal technique in constructing the canonical dynamics for such geometries. This is then used in section \ref{sec_deformation} to derive the deformation algebra of initial data surfaces in any tensorial spacetime geometry. Since geometrodynamics must evolve geometric initial data from one initial data hypersurface to the next, gravitational dynamics must represent this deformation algebra on a geometric phase space, which we explain in section \ref{sec_canonical}. 
Chapters \ref{sec_ppgrav} and \ref{sec_fieldgrav} then deal with different incarnations of the same program; the former derives the dynamics of dispersion relations independent of any underlying geometric tensor, while the latter deals with the more fundamental question of deriving dynamics for the fundamental geometric tensor.  
More specifically, the supermomentum for the dispersion relation geometry is constructed in section \ref{sec_supermomentum}, while  the corresponding superhamiltonian splits into a non-local part that we construct in section \ref{sec_nonlocal} and a local part that is determined by equations that we derive in the course of sections \ref{sec_Lagrangian}, \ref{sec_reduction} and \ref{sec_construction}. Section \ref{sec_firstderivative} then derives further insight on the structure of the equations determining the local part of the  superhamiltonian for dispersion relations, before we recover the dynamics for the dispersion relation of standard general relativity in section \ref{sec_gr}, starting from nothing more than the dispersion relation of lowest possible degree, demonstrating that the general principles underlying our study of all spacetimes are none other than those also underlying general relativity. The construction of the supermomentum and superhamiltonian determining the dynamics of a fundamental geometric tensor field in chapter \ref{sec_fieldgrav} proceeds very much along the same lines, but depends heavily on the algebraic structure of the fundamental geometric tensor one considers. Building on some work recycleable from section \ref{sec_ppgrav}, we derive the equations whose solution yields the dynamics for area metric manifolds as a prototypical example in section \ref{sec_candyn} after having constructed the relevant geometric phase space in section \ref{sec_fieldphasespace}. We conclude in section \ref{sec_conclusions}.

\newpage\section{Kinematics of tensorial spacetimes\label{sec_kinematics}}
Whether a tensor field can provide a spacetime structure depends on the matter one wishes to consider on it. In this chapter we will first review how three fundamental requirements one needs to ask of any realistic matter theory---predictivity, interpretability and quantizability---greatly restrict the tensor field backgrounds they can couple to. In particular, we will see that the dispersion relation associated with the entirety of field equations on a spacetime plays a prominent geometric role, and almost single-handedly encodes the kinematics associated with the underlying tensorial geometry. These insights, which are reviewed here in section \ref{sec_primer} in a brief but self-contained manner for the benefit of the reader, are then used in section \ref{sec_deformation} to study the deformation algebra of initial data hypersurfaces in tensorial spacetimes. The basic geometrodynamic idea, namely to use this deformation algebra to derive canonical dynamics, is then laid out in section \ref{sec_canonical} and presents the key to constructing  gravitational dynamics for the dispersion relation in chapter \ref{sec_ppgrav}, or more fundamentally, the  underlying tensorial geometry in chapter \ref{sec_fieldgrav}.

\subsection{Primer on tensorial spacetime geometries\label{sec_primer}}
In this section, we give a concise review of 
tensorial geometries that can 
serve as a spacetime structure. 
The technical proofs underlying this summary are presented in detail in \cite{RRS} and rather pedagogical fashion in the lecture notes \cite{FPSnotes}. 
To aid the reader's intuition, we illustrate each abstract construction in this section immediately for the familiar example of a standard metric geometry, 
before moving on to the next construction. Occasionally we will also contrast this to area metric geometry \cite{PSW,HehlObukhov} as a comparatively well-studied example for a non-metric tensorial geometry. Having studied the general theory and these examples, the reader should be in a position to carry out a similar analysis for his or her favourite tensorial geometry.

All we know about spacetime, we know from probing it with matter \footnote{This point has been made very lucidly, and very close to the spirit of the present paper, by Laemmerzahl in \cite{laemmerzahl}, who explores what can be learnt about the spacetime geometry from the predictivity of linear matter field equations alone.}. 
So we consider, in addition to an a priori arbitrary tensor field $G$ (the ``geometry'') on a smooth manifold $M$ also a field $\phi$ (the ``matter''), which takes values in some tensor representation space $V$ and whose gauge-fixed dynamics are encoded in linear field equations that transform as a tensor. Since the only other structure on the manifold besides the matter field $\phi$ is provided by the geometric tensor $G$, the coefficients $Q$ of the matter field equations (after removing potential gauge symmetries and separating off the related constraint equations) must be built entirely from the geometric tensor and its partial derivatives \footnote{if the linear matter field equations are obtained from a linearization of non-linear matter field equations around an exact solution $\phi_0$ of the latter, the coefficients $Q$ of the linearized equation will generically depend also on $\phi_0$},
\begin{equation}\label{mattereqn}
  \sum_{n=0}^N Q[G]_{AB}^{a_1 \dots a_n}(x) \partial_{a_1} \dots \partial_{a_n} \phi^B(x) = 0\,,
\end{equation}
with small latin spacetime indices running from $0, \dots, \dim M - 1$ and capital latin respresentation space indices ranging over $1, \dots, \dim V$.
It is straightforward to establish that in such an equation the leading order coefficient, and generically only this one, transforms as a tensor, if the entire equation does \footnote{Transformation of the entire equation as a tensor is guaranteed, for instance, if the equation is obtained by variation of a scalar action $S[G,\phi]$ with respect to the tensor field $\phi^A$---note the corresponding position of the index $A$ in Eqn. (\ref{mattereqn})}.
This will render definition (\ref{principal}) below covariant. For the example of the geometry being given by an inverse metric tensor field ($G^{ab}=G^{ba}$ with non-zero determinant everywhere) and a scalar field $\phi$ ($\dim V = 1$) obeying the massless Klein Gordon equation $G^{a_1 a_2} \partial_{a_1} \partial_{a_2} \phi - \Gamma^{a_1}{}_{mn} G^{mn} \partial_{a_1} \phi = 0$, where $\Gamma$ are the Christoffel symbols of the metric $G_{ab}$, we indeed find that the leading quadratic order coefficient transforms as a tensor, while the linear order coefficient does not (and could not, since it must ensure that the enrire equation transforms as a scalar). 


Requiring that matter equations of the form (\ref{mattereqn}) are predictive, interpretable and quantizable imposes necessary conditions on the underlying geometry $G$. These conditions have been derived and explained in detail in \cite{RRS}. Here we present a practical summary of these conditions and their implications as far as they are directly relevant for the present article. 
All constructions revolve around the totally symmetric contravariant tensor field $P$ defined from the leading order coefficients of the matter field equations (\ref{mattereqn}) by
\begin{equation}\label{principal}
   P^{i_1 \dots i_{\deg P}}(x) k_{i_1} \dots k_{i_{\deg P}} := \rho \det_{A,B}\left(Q[G]_{AB}^{a_1\dots a_N}(x) k{a_1} \dots k_{a_N}\right)
\end{equation}
for all points $x\in M$ and cotangent vectors $k\in T_x^*M$ and a scalar density function $\rho$ constructed from the geometry $G$ such as to be of opposite density weight to the determinant over the tensor representation indices. 
To lighten the notation, we will often use the shorthand $P(x,k)$ for the left hand side of Eq. (\ref{principal}). Furthermore we may agree, since no information is lost and it is technically convenient, to remove any repeated factors into which the field $P$ may factorize; so if the above construction yields $P(x,k) = P_1(x,k)^{\alpha_1} \cdots P_f(x,k)^{\alpha_f}$ then we consider instead the reduced tensor field $P$ defined by $P(x,k)=P_1(x,k) \cdots P_f(x,k)$.
The physical meaning of the tensor field $P$ is revealed by the eikonal equation \cite{Perlick} for the dynamics (\ref{mattereqn}), which shows that
\begin{equation}
  P(x,k) = 0\,
\end{equation}
is the dispersion relation that a covector $k \in T_x^*M$ must satisfy in order to qualify as a massless momentum.
For our simple example of a Klein-Gordon field on a metric geometry, the determinant in (\ref{principal}) is of weight zero, and for the choice $\rho = 1$ we obtain $P^{i_1 i_2} = G^{i_1 i_2}$, and one indeed recovers the familiar massless dispersion relation $G^{a_1a_2} k_{a_1} k_{a_2} = 0$. An instructive non-metric example is provided by abelian gauge theory coupled to an inverse area metric tensor geometry \cite{HehlObukhov,lightprop}, which is based on a fourth rank contravariant tensor field $G$ featuring the algebraic symmetries $G^{abcd}=G^{cdab}=-G^{bacd}$; calculation of the principal polynomial (after removing gauge-invariance, observing resulting constraints on initial conditions and re-covariantizing the expression) one obtains \cite{Rubilar,PSW,SWW} in $\dim M = d$ dimensions the totally symmetric tensor field
\begin{equation}
\label{PArea}
P^{i_1 \dots i_{2(d-2)}} =\rho(G)\, \epsilon_{a a_1 \dots a_{d-1}} \epsilon_{b_1\dots b_{d-1} b} G^{a a_1 b_1 (i_1} G^{i_2|a_2b_2|i_3} \dots G^{i_{2(d-3)}|a_{d-2} b_{d-2}| i_{2d-5}} G^{i_{2(d-2)})a_{d-1} b_{d-1} b}
\end{equation}
of tensor rank $\deg P = 2(d-2)$, with some scalar density $\rho(G)$ of weight $+2$ constructed from $G$. In four spacetime dimensions, for example, where the area metric may be decomposed into a cyclic part $G_C$ with $G_C^{a[bcd]}=0$ and a totally antisymmetric part given in terms of a scalar density $\Phi$ of weight $-1$, $G^{abcd}=G_C^{abcd}+\Phi\epsilon^{abcd}$, one may chose $\rho(G)=-1/(24\Phi^2)$. This non-trivial example for a field $P$ illustrates two salient points. First, it reveals what a dramatic accident it is that in Lorentzian geometry the field $P$, which will be central to all further developments, is essentially identical to the fundamental geometric field $g$; for in area metric geometry, one not only sees that $P$ is a tensor field of generically entirely different tensor rank than the underlying fundamental geometric tensor $G$, but may also feature an entirely different index symmetry structure: the tensor $P$ is always totally symmetric. Second, it exemplifies the rule that generically the fundamental geometry $G$ cannot be reconstructed from the field $P$ \footnote{since not even all degrees of freedom in the area metric $G$ enter the field $P$, see e.g. \cite{Rubilar}}. 
With these remarks on the role of the field $P$ as a dispersion relation and its origin in matter field equations coupled to some tensorial geometry, we are now prepared to lay down the three crucial algebraic conditions that the tensor field $P$ needs to satisfy. These conditions in turn restrict the geometric tensor $G$ that underlies $P$ \footnote{This dependence from specific matter equations is not a weakness of the approach, but rather an insight: there are no viable or non-viable spacetime structures as such, but they prove their validity by enabling specific field equations to be predictive. The causality of Lorentzian manifolds is in fact the causality of Maxwell theory, and thus of all the other matter dynamics modelled in the image of Maxwell theory.}. 

The first condition, predictivity of the matter field equations, translates into the algebraic requirement that the tensor field $P$ be hyperbolic \cite{Garding,laemmerzahl}. This means that there exists a covector field $h$ with $P(h)>0$ such that for all covector fields $r$ there are only real functions $\lambda$ on $M$ such that
\begin{equation}
   P(x,r(x)+\lambda(x)h(x)) = 0
\end{equation}
everywhere. Obviously if $P(h)<0$, one could arrange for $P(h)>0$ simply by changing the overall sign of the density $\rho$ appearing in (\ref{principal}), and we will agree to do so for definiteness \footnote{This choice is the general tensorial analogue to a choice of `mainly minus' signature $(+-\dots-)$ in the special case of metric geometry}.
In any case, it is useful terminology to call a covector field $h$, if it indeed exists, a hyperbolic covector field with respect to $P$. Only hypersurfaces whose canonical normal covector fields (defined to annihilate any tangent vector field to the hypersurface) are hyperbolic can serve as initial data surfaces for equations of the type (\ref{mattereqn}). We will return to this point later.  
For our example of a metric geometry, it is easy to check that $P^{i_1i_2}=G^{i_1i_2}$ is hyperbolic if and only if the inverse metric has Lorentzian signature $(+-\dots-)$, and that the hyperbolic covectors are exactly those covectors for which $P^{i_1i_2}h_{i_1}h_{i_2}>0$; in other words, initial data surfaces need to be spacelike. The reader be warned, however, that such a simple characterization of hyperbolic covectors and thus initial data surfaces merely by the sign of their co-normals or tangent vectors under $P$ is not generic and merely a coincidence in the metric case. The underlying general definitions however work for all geometries.

The second condition, interpretability of the matter field equations translates into a time-orientability condition for the underlying geometry. This is simply the algebraic requirement that also the so-called dual tensor field $P^\#$ be hyperbolic. 
Indeed, for any hyperbolic tensor field $P$, one can show that there always exists a totally symmetric covariant dual tensor field $P^\#$ of some rank $\deg P^\#$ such that
\begin{equation}
  P^\#(x, DP(x,k(x))) = 0 
\end{equation} 
for all covector fields $k$ with $P(x,k(x))=0$ everywhere, where 
$DP(x,q)$ denotes the vector with components $DP(x,k)^a=(\deg P) P^{a f_2\dots f_{\deg P}} k_{f_2} \dots k_{f_p}$ and we used a shorthand for the evaluation of the field $P^\#$ on a vector that is analogous to the previous one for $P$ on a covector. The dual tensor field $P^\#$ is unique up to a real conformal factor, and can always be constructed, essentially by determining a Gr\"obner basis \cite{Hassett}. For our example of metric geometry, a dual of $P^{i_2i_2}=G^{i_1i_2}$ is given by $P^\#_{i_1i_2} = G_{i_1i_2}$, as one easily verifies. Returning to the general case, time-orientability means that there exists a vector field $H$ such that for every vector field $R$ there are only real functions $\mu$ on $M$ such that 
\begin{equation}
  P^\#(x, R(x)+\mu(x)H(x) ) = 0
\end{equation} 
everywhere. A vector field $H$ satisfying this condition will be called a time-orientation. Once a time-orientation has been chosen, it is useful to consider, separately in each tangent space, the connected set $C^\#_x$ of all hyperbolic vectors to which the vector $H_x$ of the time-orientation belongs.
According to a classic theorem \cite{Garding}, $C^\#_x$ constitutes an open and convex cone in the tangent space $T_xM$, and we will call $C^\#_x$ the cone of observer tangents (to observer worldlines through the point $x$). Note that in general, hyperbolicity of $P$ does not already imply hyperbolicity of $P^\#$, and thus predictivity does not imply time-orientability in general. For our metric example, however, it trivially does; $P^\#_{i_1i_2}=G_{i_1i_2}$ is hyperbolic if and only if $P^{i_1i_2}=G^{i_1i_2}$ is, and the cones $C^\#$ of observer tangents are the timelike vectors $X$ at each point that are future-oriented with respect to some global timelike vector field $T$ representing the time-orientation, i.e., satisfy $G_{a_1a_2} X^{a_1} T^{a_2} > 0$; again, this simple sign condition to decide membership of $X$ in the cone of observer tangents selected by $T$ is a coincidence in the metric case, and again has to be replaced by the underlying general definition above for other geometries.

The third condition on the matter field equations, namely that these are quantizable, is that the geometry be energy-distinguishing. 
This simply means that all observers agree on the sign of the energy of a massless momentum. More precisely, a geometry is energy-distinguishing if for every point $x\in M$ and every massless momentum $k$ either $k(X)>0$ or $-k(X)>0$ for all $X \in C^\#_x$. In a hyperbolic, time-orientable and energy-distinguishing geometry, one can then also meaningfully define {\sl massive} particle momenta of positive energy at some point $x$ as those hyperbolic covectors $q \in T_x^*M$ for which $q(X)>0$ for all observers $X \in C^\#_x$. 
To be able to do this is of crucial importance when performing a split of a basis of solutions to the field equations into positive and negative frequency solutions in a canonical quantization of the matter field. 
These massive positive energy momenta constitute an open convex cone $C_x$ in the cotangent space at $x$. The mass $m$ of such a positive energy massive particle momentum $q\in C_x$ is then defined by 
\begin{equation}\label{massive}
  P(x,q) = m^{\deg P}\,.
\end{equation}
In Lorentzian metric geometry, the above definitions of course recover as the  positive energy massive and massless momenta precisely those timelike and null covectors whose application to a future-directed timelike vector is positive. It may be worth emphasizing again that for a covector to qualify as massive, it must not only satisfy the massive dispersion relation (\ref{massive}) but indeed be hyperbolic, as stipulated above. Only in Lorentzian geometry does the massive dispersion relation already imply hyperbolicity.

Only if a geometry satisfies the three conditions laid out above can one associate worldlines with the massless and massive dispersion relations. For only then can one solve for the momenta $q$ after variation of the Hamiltonian actions
\begin{equation}
  S_{\textrm{massless}}[x,q,\lambda] = \int d\tau\, \left[\dot x^a q_a - \lambda P(q)\right]\quad \textrm{ and } \quad S_{\textrm{massive}}[x,q,\lambda] = \int d\tau\,\left[\dot x^a q_a - \lambda \ln P(\frac{q}{m})\right]\,,
\end{equation}
respectively. Defining the Legendre map $L_x$ for all covectors $q$ in the open convex cone $C_x$ of positive energy massive momenta at some point $x$ as
\begin{equation}
  L^a(x,q) = \frac{P(x)^{a b_2 \dots b_{\deg P}} q_{b_2} \dots q_{b_{\deg P}}}{P(x,q)}\,,
\end{equation}
which by virtue of the energy-orientability of $P$ possesses a unique inverse $L^{-1}_x$ on its domain, one finds \cite{RRS} that the worldlines of free massless and massive particles are stationary curves of the reparametrization-invariant Lagrangian actions
\begin{equation}
  S_\textrm{massless}[\mu,x] = \int d\tau\, \mu P^\#(\dot x) \qquad \textrm{ and } \qquad S_\textrm{massive}[x] = m\int d\tau \, P(L^{-1}(\dot x))^{-\frac{1}{\deg P}}\,,
\end{equation}
respectively. The massive particle action reveals the physical meaning of the Legendre map $L$, since one readily derives that the canonical momentum of a positive energy massive particle is related to the worldline tangent vector as $q = m L^{-1}(\dot x)$ if one chooses the proper time parametrisation $P(L^{-1}(\dot x))=1$ along the worldline. Put simply, the Legendre map raises the index on a positive energy massive momentum, in one-to-one but non-linear fashion. For the example of Lorentzian geometry, we find that under the familiar proper time parametrisation $G_{ab}\dot x^a\dot x^b=1$, the worldline tangent vector $\dot x$ and the corresponding particle momentum $q$ of a particle of mass $m$ are related through $m\dot x^a=G^{ab}q_b$. The massless and massive Lagrangian actions for the free point particle on a Lorentzian spacetime recover the standard textbook postulates.

Of central importance for the aim of this article, namely to derive the dynamics of hyperbolic, time-orientable and energy-orientable geometries, is the following insight. Hypersurfaces that are potential carriers of initial data and at the same time accessible by observers are those whose co-normal at each point lie in the cone $L^{-1}(C^\#)$. This is because, on the one hand,
the purely spatial directions seen by an observer with tangent vector $X\in C^\#$ are precisely those vectors annihilated by the covector $L^{-1}(X)$. On the other hand, the cone of these observer co-tangents can be shown to always lie within the cone $C$ of hyperbolic covectors for hyperbolic, time-orientable and energy-distinguishing geometries. In Lorentzian geometry, such inital data surfaces accessible to observers are simply the spacelike hypersurfaces. Incidentally, only when $L^{-1}(C^\#)$ does not only lie within $C$, but entirely coincides with it, is the theory free of particles travelling faster than the speed of some light \footnote{If there are such superluminal particles, they can radiate off massless particles until they are infraluminal, see \cite{FPSnotes}}. 

The deformation of such observer-accessible initial data hypersurfaces, separately in normal and tangential directions, will be the topic of the next section. While a generic hypersurface directly gives rise to tangent directions, but merely normal co-directions, it is only the Legendre map (and thus the spacetime geometry) that allows to associate a normal co-direction $n$ of a hypersurface with a corresponding normal direction $T=L(n)$ if $n$ lies in $L^{-1}(C^\#)$. Normalizing the latter by requiring $P(L^{-1}(T))=1$ corresponds to requiring that the normal direction be tangent to an observer worldline with proper time parametrization.  Thus the normal deformation of observer-accessible initial data hypersurfaces feels the spacetime geometry
 through the Legendre map. It is this role of the Legendre map that we will see to hold the key to the derivation of the gravitational dynamics for general tensorial spacetimes.

\subsection{Deformation of initial data surfaces\label{sec_deformation}}
The aim of this paper is to find dynamics that develop initial geometric data from one initial data hypersurface to another, such that sweeping out the spacetime manifold in this way one reconstructs a hyperbolic, time-orientable and energy-distinguishing dispersion relation everywhere. In this section we describe initial data hypersurfaces by their embedding maps and study how functionals of this embedding map change under normal and tangential deformations of the hypersurface. The functionals of interest later on will be the induced geometry seen by point particles in section \ref{sec_ppgrav} or the induced geometry seen by fields in section \ref{sec_fieldgrav}. The change of a generic functional of the embedding map can be expressed by a linear action of deformation operators on such functionals, and it is the commutation algebra of these deformation operators that we are after \footnote{Employing the techniques for general tensorial spacetime geometries instead of those valid only for the special case of Lorentzian manifolds, this section follows closely the philosophy and calculation of \cite{HKT}}.

More precisely, we consider a hypersurface $X(\Sigma)$ defined by an embedding map $X: \Sigma \hookrightarrow M$ of a smooth manifold $\Sigma$ with local coordinates $\{y^\alpha\}$ into the smooth manifold $M$ with local coordinates $\{x^a\}$; here and in the remainder of this paper, latin `spacetime' indices run from $0$ to $\dim M\min1$ while greek `hypersurface' indices run from $1$ to $\dim M\min1$. Without additional structure, the embedding defines at each point $y$ of the hypersurface $\dim M\min1$ spacetime vectors
\begin{equation}
  e_\alpha(y) = \frac{\partial X^a(y)}{\partial y^\alpha} \frac{\partial}{\partial x^a}
\end{equation}
tangent to the hypersurface $X(\Sigma)$, which in turn define, up to scale, normal spacetime covectors $n(y)$ as the annihilators of all tangent vectors,
\begin{equation}
   n(y)(e_{\alpha}(y)) = 0 \qquad \alpha = 1, \dots, \dim M\min1\,.
\end{equation}
Only if we restrict attention to initial data hypersurfaces whose data are accessible to observers, by requiring that the $n(y)$ lie in the respective cones $L^{-1}(C^\#)$ everywhere along the hypersurface $X(\Sigma)$, can we impose the normalization $P(n(y))=1$ and thus obtain a unique spactime vector field $T(y)=L(n(y))$ representing the normal directions, rather than normal co-directions, away from the hypersurface. Thus an accessible initial data hypersurface $X(\Sigma)$ induces a complete spacetime tangent space basis $\{T(y), e_1(y), \dots, e_{\dim M\min1}(y)\}$ at every of its points, and dual basis 
$\{n(y), \epsilon^1(y), \dots, \epsilon^{\dim M\min1}(y)\}$ in cotangent space. 

We now consider deformations of the hypersurface $X(\Sigma)$. Technically, this is done by prescribing a smooth one-parameter family $X_t$ of embedding maps such that the original embedding map $X$ is recovered for $t=0$. Then the connecting vector field $\partial X_t/\partial t$ in spacetime, between the hypersurfaces $X_t(\Sigma)$ of this family, can be uniquely decomposed along the undeformed hypersurface into a sum of a purely spatial and a purely tangential part,
\begin{equation}\label{deformation}
  \dot X(y) = N(y)\, T^a(y) + N^\alpha(y)\, e^a_\alpha(y)\,,
\end{equation}
where the hypersurface scalar field $N$ and hypersurface vector field components $N^\alpha$ are given by
\begin{equation}
  N(y) = n(y)(\dot X(y))\quad \textrm{ and } \quad N^\alpha(y)=\epsilon^\alpha(y)(\dot X)\,
\end{equation}
and thus completely parametrize any small deformation of the embedding map $X$ into $X + dt \dot X$.

The linear change of functionals under changes of the embedding map is conveniently studied in terms of normal and tangential deformation operators. More precisely, we define the normal deformation operator 
\begin{equation}\label{Hoperator}
  \mathcal{H}(N) = \int_\Sigma dy\, N(y) T^a(y) \frac{\delta}{\delta X^a(y)}\,,
\end{equation}
acting on arbitrary functionals $F$ of the embedding function. The change of such $F$ under the deformation (\ref{deformation}) is then given to first order by $\mathcal{H}(N) F$. Similarly one obtains for a purely tangential deformation the first order change $\mathcal{D}(N^\alpha\partial_\alpha) F$ through the tangential deformation operator
\begin{equation}\label{Doperator}
  \mathcal{D}(N^\alpha \partial_\alpha) = \int_\Sigma dy\, N^\alpha(y) e^a_\alpha(y) \frac{\delta}{\delta X^a(y)}\,.
\end{equation}
A trivial check on the geometric meaning, which this terminology attaches to these operators, is their action on the components of the embedding map itself; with the definitions of the delta distribution and functional derivatives one finds
\begin{equation}
  \mathcal{H}(N) X^a(z) = N(z) T^a(z) \qquad \textrm{ and } \qquad \mathcal{D}(N^\alpha\partial_\alpha) X^a(z) = N^\alpha(z) e^a_\alpha(z)\,,
\end{equation}
which indeed are precisely the normal and tangential components of the deformation (\ref{deformation}). Since the embedding is a linear functional of itself, this shows that (\ref{Hoperator}) and (\ref{Doperator}) indeed are the operators that bring about the normal and tangential deformations of functionals to linear order, as desired. 

Finally we may calculate their commutator algebra. The latter will play a crucial role througout this paper. Now the basis vectors $T$ and $e_1, \dots, e_{\dim M\min1}$ are functionals of the embedding map, and it is thus clear that multiple application of deformation operators will require to know their functional derivatives with respect to the embedding functions. While for the hypersurface tangent vectors one obtains 
\begin{equation}
  \frac{\delta e^a_\alpha(y)}{\delta X^b(z)} = - \delta^a_b \partial_\alpha \delta_y(z)
\end{equation}
in straightforward fashion directly from their definition, one needs to work somewhat harder from the definition of $T$ to find
\begin{eqnarray}
  \frac{\delta T^a(y)}{\delta X^b(z)} &=& (\deg P\min1)(e^a_\alpha n_b P^{\alpha\beta})(y)\, \partial_\beta\delta_y(z)\nonumber\\
& & +
 \left[n_{j_2}\dots n_{j_{\deg P}}\partial_bP^{a j_2\dots j_{\deg P}}-\frac{\deg P\min 1}{\deg P}T^a n_{j_1}\dots n_{j_{\deg P}}\partial_b P^{j_1\dots j_{\deg P}}\right](y)\delta_y(z),
\end{eqnarray}
where in the first summand one of the hypersurface tensors defined in (\ref{hypgeo}) appears. Note that the dispersion relation enters only in the variation of the normal vector, but not of the tangent vectors. This is because the definition of the former employs the Legendre map defined by the dispersion relation.
This is indeed the way the geometry enters into the deformation algebra, which is now straightforwardly calculated to be \footnote{The minus sign in the algebra equation involving two normal deformation operators is due to our normalisation condition $P(n)=1$. In most standard texts on the hypersurface deformation algebra of Lorentzian manifolds the normalisation is chosen to be $P(n)=-1$, which results in a plus sign in the first algebra equation.}
\begin{eqnarray}
  {[\mathcal{H}(N),\mathcal{H}(M)]} &=& -\mathcal{D}((\deg P\min1) P^{\alpha\beta}(M \partial_\beta N - N \partial_\beta M)\partial_\alpha)\,,\label{HH}\\
  {[\mathcal{D}(N^\alpha \partial_\alpha),\mathcal{H}(M)]} &=& - \mathcal{H}(N^\alpha \partial_\alpha M)\label{DH}\,,\\
  {[\mathcal{D}(N^\alpha\partial_\alpha),\mathcal{D}(M^\beta \partial_\beta)]} &=& - \mathcal{D}((N^\beta \partial_\beta M^\alpha - M^\beta \partial_\beta N^\alpha)\partial_\alpha)\,.\label{DD}
 \end{eqnarray}
The exclusive appearance of the hypersurface tensor field components $P^{\alpha\beta}$ induced from the spacetime tensor  $P$ by virtue of $P^{\alpha\beta}=P^{abf_1\dots f_{\deg P - 2}} \epsilon^\alpha_a \epsilon^\beta_b n_{f_1} \dots n_{f_{\deg P -2}}$ on the right hand side of the commutator of two normal deformation operators---the last two commutators are indeed fully independent of the hypersurface geometry---originates entirely in the use of the Legendre map when defining the spatial fields. Thus $P^{\alpha\beta}$ appears irrespective of which type of geometry on the hypersurfaces one chooses to study (possible choices are the pullbacks of the geometry seen by point particles considered in chapter \ref{sec_ppgrav} or the geometry seen by fields considered in chapter \ref{sec_fieldgrav}). 
The calculation of the $P^{\alpha\beta}$ on the right hand side of the algebra above in terms of the hypersurface geometry is just more complicated for geometries seen by fields than for geometries seen only by point particles, but it is always the $P^{\alpha\beta}$ that appears there. 
Finally it is useful to observe that  the tangential deformation operators constitute a subalgebra. 

\subsection{Towards canonical dynamics for hypersurface geometries\label{sec_canonical}}
So far in this paper, we tacitly assumed to have knowledge about the values of the geometric tensor $G$---and thus also the cotangent bundle function $P$ derived from it---at every point of the entire spacetime manifold $M$. This enabled us to derive how any functional $F$ of a hypersurface embedding map $X: \Sigma \hookrightarrow M$ changes under a change (\ref{deformation}) of the embedding map. We are particularly interested in the particular type of functionals of the embedding map that arise as  normal and tangential projections of a spacetime $(r,s)$-tensor field $F$ to an embedded hypersurface. For simplicity, consider a $(1,0)$-tensor field $F$ on M, which induces the projections
\begin{equation}
  F^0(y)[X] := F(n(y)) \qquad \textrm{ and } \qquad F^\alpha(y)[X] := F(\epsilon^\alpha(y)) \quad\textrm{ for } \alpha=1,\dots,\dim M-1\,,
\end{equation}
which yields the collection of functionals $F^A(y)=(F^0(y),F^\alpha(y))$, where we used the spacetime covector frame $\{n,\epsilon^1, \dots \epsilon^{\dim M -1}\}$ along the hypersurface to project $F$. One proceeds analogously for tensor fields of valence $(r,s)$. Knowing the value of the tensor field $F$ throughout spacetime, and in particular in a neighborhood of an embedded hypersurface $X(\Sigma)$, we can write the linear change of the functionals $F^A(y)$ under a deformation of the original hypersurface controlled by the lapse $N$ and shift $N^\alpha$ as
\begin{equation}\label{hypsurfdef}
  \int_\Sigma dz \left[N(z) \mathcal{H}(z) + N^\alpha(z) \mathcal{D}_\alpha(z)\right] F^A(y)[X]\,,
\end{equation}  
where we introduced the localized operators $\mathcal{H}(z):=\mathcal{H}(\delta_z)$ and $\mathcal{D}_\alpha(z):=\mathcal{D}(\delta_z \partial_\alpha)$.

But this omniscient view of the values of the tensor field $F$ and the geometry $G$ at every point of spacetime is not afforded by us mere mortals. What we have access to, at best, are the values  $\hat{F}^A(y)$ on $\Sigma$, understood as mere hypersurface tensor fields, rather than functionals of the embedding map. If we then wish to predict the values of the $\hat{F}^A$ on some different hypersurface through spacetime, we need to stipulate how these fields change from the initial hypersurface $X(\Sigma)$ to a deformed one near-by, and we will see in a moment that we are rather constrained in the way we can stipulate such equations of motion. Anyway, since we are ignorant of any of the field values of $F$ away from the hypersurface, we need to compensate for this lack of knowledge by adjoining canonical momentum densities $\hat{\phi}_A$ of weight one to each configuration variable $\hat F^A$, which is equivalent to introducing a Poisson bracket
\begin{equation}
  \big\{\hat C, \hat D\big\} := \int_\Sigma dz \left[\frac{\delta \hat{C}}{\delta \hat{F^A}} \frac{\delta \hat{D}}{\delta \hat{\phi}_A} - \frac{\delta \hat{D}}{\delta \hat{F^A}} \frac{\delta \hat{C}}{\delta \hat{\phi}_A}\right]
\end{equation}
on the space of functionals of the phase space variables $(\hat F^A,\hat \phi_A)$, which is sometimes referred to as superspace. One can then give dynamics to the hypersurface fields $\hat F^A$ by stipulating that their values change by the amount
\begin{equation}\label{classHam}
  \big\{\hat{F}^A(y), \int_\Sigma dz \left[N(z) \hat{\mathcal{H}}(z) +  N^\alpha(z) \hat{\mathcal{D}}_\alpha(z) \right] \big\}
\end{equation}
when evolved to a neighboring hypersurface whose deformation from the initial one is determined by the lapse $N$ and shift $N^\alpha$, where the quantities $\hat{\mathcal{H}}$ and $\hat{\mathcal{D}}_\alpha$ are some a priori arbitrary functionals of the phase space variables $(\hat{F}^A,\hat{\phi}_A)$. 
For brevity, and in accordance with the standard terminology in geometrodynamics, we will refer to $\hat{\mathcal{H}}(y)$ as the superhamiltonian and to $\hat{\mathcal{D}}_\alpha(y)$ as the supermomentum. The dynamics (\ref{classHam}) are further assumed to be supplemented by first class constraints  
\begin{equation}\label{constraints}
  \hat{\mathcal{H}}(y) \approx 0 \qquad \textrm{ and } \qquad \hat{\mathcal{D}}_\alpha(y) \approx 0
\end{equation}
implementing the required diffeomorphism gauge symmetry. 

It is clear that if the dynamically evolved hypersurface field values on the deformed hypersurface are to coincide with what the hypersurface deformation (\ref{hypsurfdef}) yields, independent of any particular deformation $(N,N^\alpha)$, then we must require that (\ref{hypsurfdef}) coincides with (\ref{classHam}), or equivalently,
\begin{equation}\label{application}
  \mathcal{H}(y) F^A(y)[X] = \big\{\hat{F}^A(y), \hat{\mathcal{H}}(z)(y)\big\} \qquad \textrm{ and } \qquad  \mathcal{D}_\alpha(y) F^A(y)[X] = \big\{\hat{F}^A(y), \hat{\mathcal{D}}_\alpha(z)(y)\big\}\,.
\end{equation} 
We cannot extend these equations to the momentum variables, since we do not know at this stage how the latter can be understood as functionals of the embedding map---this is for the dynamics to determine. But using the relations (\ref{application}) in the deformation algebra (\ref{HH}), (\ref{DH}), (\ref{DD}) and the Jacobi identity for the Poisson bracket, one finds that a {\it sufficient} condition for our above compatibility requirement is that the 
functionals
\begin{equation}\label{phasespaceHD}
  \hat{\mathcal{H}}(N) := \int_\Sigma dy\, N(y) \hat{\mathcal{H}}(y) \qquad \textrm{ and } \qquad \hat{\mathcal{D}}(N^\alpha \partial_\alpha) := \int_\Sigma dy\, N^\alpha(y) \hat{\mathcal{D}}_\alpha(y)
\end{equation}
of the phase space variables have Poisson brackets that represent the deformation algebra commutation relations \footnote{The change of the overall sign on the right hand side of the Poisson algebra equations in comparison to the hypersurface deformation algebra is due to the action from the left of the deformation operators on hypersurface functionals as oppposed to the insertion of the supermomentum and superhamiltonian in the right slot of the Poisson bracket when acting on a phase space functional.}
\begin{eqnarray}
  \{\hat{\mathcal{H}}(N),\hat{\mathcal{H}}(M)\} &=& \hat{\mathcal{D}}((\deg P\min1) \hat{P}^{\alpha\beta}(M \partial_\beta N - N \partial_\beta M)\partial_\alpha)\,,\label{repHH}\\
  \{\hat{\mathcal{D}}(N^\alpha \partial_\alpha),\hat{\mathcal{H}}(M)\} &=&  \hat{\mathcal{H}}(N^\alpha \partial_\alpha M)\,,\label{repDH}\\
  \{\hat{\mathcal{D}}(N^\alpha\partial_\alpha),\hat{\mathcal{D}}(M^\beta \partial_\beta)\} &=&  \hat{\mathcal{D}}((N^\beta \partial_\beta M^\alpha - M^\beta \partial_\beta N^\alpha)\partial_\alpha)\,.\label{repDD}
 \end{eqnarray}
The extent to which this representation requirement is not {\it necessary} to satisfy our compatibility condition, however, precisely encodes the information concerning the functional dependence of $\hat{\mathcal{H}}$ and $\hat{\mathcal{D}}$ on $\hat F^A$, while (\ref{application}) already determines $\delta\hat{\mathcal{H}}(z)/\delta \hat\phi_A(y)$ and $\delta\hat{\mathcal{D}}_\alpha(z)/\delta \hat\phi_A(y)$. We will return to this point when constructing $\hat{\mathcal{H}}$ and $\hat{\mathcal{D}}$ from the above algebra, and indeed the major part of the remainder of this paper will be devoted to this construction. 

At this point the paper splits into two different projects. While both are concerned with finding gravitational dynamics---by way of finding the supermomentum $\hat{\mathcal{D}}_\alpha$ and $\hat{\mathcal{H}}$ satisfying the Poisson algebra (\ref{repHH}),(\ref{repDH}),(\ref{repDD}) according to the programme laid out in this section---they do so for different geometric degrees of freedom. The first project, treated in chapter \ref{sec_ppgrav}, considers the projections of the components of the dispersion relation $P$ as the geometric degrees of freedom on a hypersurface, and derives the equations that determine its superhamiltonian and supermomentum. In contrast, the second project, treated in chapter \ref{sec_fieldgrav}, illustrates how the projections of the fundamental geometric tensor $G$ are taken as the degrees of freedom to which one gives dynamics. This point of view is more fundamental, but its details depend heavily on the algebraic properties of the geometric tensor.

\newpage\section{Dynamics of dispersion relations\label{sec_ppgrav}}
Point particles only see those aspects of a tensorial spacetime geometry $G$ that are encoded in the tensor field $P$ underlying the massive and massless dispersion relations. It is a mere coincidence in Lorentzian geometry that the tensor field $P^{a_1a_2} = G^{a_1a_2}$ contains precisely the same information as the fundamental geometric tensor field $G_{ab}$ to which fields couple. For any other tensorial spacetime geometry, the totally symmetric tensor field $P$ can be expressed in terms of the fundamental geometric tensor field $G$, but not vice versa. 
So if one is interested in a full gravitational theory to which fields and point particles can couple, one needs to derive dynamics for the fundamental geometric tensor $G$, and we will do so in chapter \ref{sec_fieldgrav}. 
But if one is only interested in gravitational fields acting on, and generated by, point particles, one may alternatively construct gravitational dynamics directly for any hyperbolic, time-orientable and energy-distinguishing tensor field $P$. 
Indeed, we obtain a rather sweeping result in this chapter: we derive a system of homogenous linear partial differential equations whose solutions yield all possible canonical dynamics for physical dispersion relations.

\subsection{Phase space for geometries seen by point particles}
Since we wish to study the spatial geometry seen by point particles on an initial data hypersurface $X(\Sigma)$, we are interested in the hypersurface tensor fields $P$, $P^{\alpha_1}$, \dots, $P^{\alpha_1 \alpha_2 \dots \alpha_{\deg P}}$ that arise as functionals of the embedding map through
\begin{equation}\label{hypgeo}
  P^{\alpha_1 \dots \alpha_I}(y)[X] := P(\underbrace{\epsilon^{\alpha_1}(y), \dots, \epsilon^{\alpha_I}(y)}_{I},\underbrace{n(y), \dots, n(y)}_{\deg P\min I})\qquad \textrm{for } I=0,\dots,\deg P\,,
\end{equation}
where we used the complete spacetime covector frame $\{n,\epsilon^1,\dots,\epsilon^{\deg M -1}\}$ along the hypersurface to project the spacetime tensor field $P^{a_1\dots a_{\deg P}}$ onto $\Sigma$. The normalisation conditions $P(n)=1$ and $T^{a}\epsilon^\alpha_a=0$ immediately imply that the two functionals $P$ and $P^{\alpha}$ are constant along $\Sigma$. This property is conserved under hypersurface deformations and thus the $P$ and $P^{\alpha}$ do not carry any dynamical information and can be discarded as configuration variables. The remaining hypersurface tensor fields however allow, in combination with the dual basis, the complete reconstruction of the spacetime dispersion relation at every point of the hypersurface and thus, in their entirety, present the geometry on the hypersurface---as it is seen by {\sl point particles}. 

As we have laid out in section \ref{sec_canonical}, the point of dynamics is to generate, not assume, the values of $P$ throughout the spacetime manifold, starting only from initial data in form of totally symmetric tensor fields
\begin{equation}\label{configvars}
  \hat{P}^{\alpha_1\alpha_2}, \quad \hat{P}^{\alpha_1\alpha_2\alpha_3}, \quad\dots,\quad \hat{P}^{\alpha_1\alpha_2 \dots \alpha_{\deg P}}
\end{equation}
on a manifold $\Sigma$ of dimension $\dim M\min1$. To lighten the notation, we collect the fields (\ref{configvars}) in a quantity $\hat{P}^A$ carrying a multi-index $A=(\alpha_1\alpha_2;\, \alpha_1\alpha_2\alpha_3;\, \dots;\, \alpha_1\alpha_2\dots\alpha_{\deg P})$ consisting of totally symmetric tuples of spacetime indices.
Starting from these configuration variables, which correspond to initial data on only one hypersurface $X_0(\Sigma)$ say, the lack of knowledge about the value of the configuration variables on neighboring hypersurfaces $X_t(\Sigma)$ must be compensated for by adjoining canonical momenta
\begin{equation}\label{momentumvars}
  \hat{\pi}_{\alpha_1\alpha_2},\quad \hat{\pi}_{\alpha_1\alpha_2\alpha_3}, \quad \dots, \quad \hat{\pi}_{\alpha_1 \alpha_2 \dots \alpha_{\deg P}}\,,
\end{equation}
or $\hat \pi_A$ for short, to the configuration variables (\ref{configvars}) on $\Sigma$. This introduction of canonical momenta is of course equivalent to adopting a Poisson bracket 
\begin{equation}\label{Poissonbracket}
  \{\hat F, \hat G\} = \int_\Sigma dy \, \left[\frac{\delta\hat F}{\delta \hat{P}^A(y)}\frac{\delta \hat G}{\delta \hat{\pi}_A(y)} - \frac{\delta\hat G}{\delta \hat{P}^A(y)}\frac{\delta \hat F}{\delta \hat{\pi}_A(y)}\right]
\end{equation}
on the space of functionals of $\hat{P}^A$ and $\hat{\pi}_A$, whose elements we denote with a hat. The configuration variables $\hat{P}^A$ and momenta $\hat{\pi}_A$ are trivally functionals of themselves, and this is the reason why we made them carry a hat from the beginning. 
 In order for the integral (\ref{Poissonbracket}) to be well-defined under changes of chart on $\Sigma$, the momenta must be tensor densities of weight one. This ultimately follows from the definition of the functional derivative. 
We now set out to determine the supermomentum $\hat{\mathcal{D}}$ and the superhamiltonian $\hat{\mathcal{H}}$ that close according to (\ref{repHH}), (\ref{repDH}) and (\ref{repDD}) and evolve the canonical variables $(\hat P^A,\hat \pi_A)$.
 
\subsection{Construction of the supermomentum}\label{sec_supermomentum}
The strategy to determine the superhamiltonian and supermomentum from the Poisson algebra---in accordance with the philosophy laid out in section \ref{sec_canonical}---follows from the fact that the supermomentum functionals constitute a subalgebra that must be solved without recourse to the further relations. This is what we do in this section. More precisely, using the definition (\ref{Doperator}) of the tangential deformation operator we may calculate the change of the functionals (\ref{hypgeo}) under tangential deformations. In order to do that, we need to know the functional derivatives of the dual hypersurface basis vectors $\{n, \epsilon^{\alpha}\}$ with respect to the embbeding map. For the hypersurface conormal, we have
\begin{equation}
\frac{\delta n_a(y)}{\delta X^b(z)}=-\frac{1}{\deg P}(n_an_{j_1}\dots n_{j_{\deg P}}\partial_b P^{j_1\dots j_{\deg P}})(y)\delta_y(z)+n_b(y)\epsilon^{\alpha}_a(y)\partial_{\alpha}\delta_y(z),
\end{equation}
while the functional derivatives of the $\epsilon^{\alpha}$ read
\begin{eqnarray}
\frac{\delta \epsilon^{\alpha}_a(y)}{\delta X^b(z)}&=&\epsilon^{\alpha}_b(y)\epsilon^{\beta}_a(y)\partial_{\beta}\delta_y(z)-(\deg P\min 1)n_a(y)n_b(y)P^{\alpha\beta}(y)\partial_{\beta}\delta_y(z)\nonumber\\
&&-\,(n_a\epsilon^{\alpha}_{j_1}n_{j_2}\dots n_{j_{\deg P}}\partial_b P^{j_1\dots j_{\deg P}})(y)\delta_y(z)\,.
\end{eqnarray}
Using these results on the left hand side in the second equation of (\ref{application}), and the second weak equality in (\ref{constraints}), one calculates that
\begin{equation}\label{PonD}
   \left\{\hat{P}^A(y),\,\hat{\mathcal{D}}(N^\alpha\partial_\alpha)\right\} =  (\mathcal{L}_{\vec{N}} \hat{P})^A(y)\,.    
\end{equation}
This is entirely plausible, since the deformation operator, and thus its represention in form of the supermomentum, push the configuration variable along the hypersurface vector field $\vec{N}$, and this is precisely what the Lie derivative is defined to do. But then it follows from the Jacobi identity for the Poisson bracket (\ref{Poissonbracket}), followed by a functional integration with respect to $\hat{\pi}$, that 
\begin{equation}\label{Ponpi}
  \left\{\hat{\pi}_A(y),\,\hat{\mathcal{D}}(N^\alpha\partial_\alpha)\right\} =  (\mathcal{L}_{\vec{N}} \hat{\pi})_A(y)\,,    
\end{equation}
for the covariant tensor densities $\pi_A$. Again this is more than plausible, since the tangential deformation operator merely reshuffles all the initial data. In summary, we obtain a set of pairwise coupled functional differential equations for all $I = 2, \dots, \deg P$, namely
\begin{eqnarray}
  \frac{\delta \hat{\mathcal{D}}(\vec{N})}{\delta \hat{P}^{\alpha_1 \dots \alpha_I}(y)} &=& (-\partial_\beta N^\beta \hat{\pi}_{\alpha_1 \dots \alpha_I} - N^\beta\partial_\beta\hat{\pi}_{\alpha_1\dots \alpha_I} - I \, \partial_{(\alpha_1} N^\beta \hat{\pi}_{\alpha_2 \dots \alpha_I) \beta})(y)\,,\\
   \frac{\delta \hat{\mathcal{D}}(\vec{N})}{\delta \hat{\pi}_{\alpha_1 \dots \alpha_I}(y)} &=& (N^\beta \partial_\beta \hat{P}^{\alpha_1 \dots \alpha_I} - I\, \partial_\beta N^{(\alpha_1} \hat{P}^{\alpha_2 \dots \alpha_I) \beta })(y)\,,
\end{eqnarray}
which are integrable since all second functional derivatives of $\hat{\mathcal{D}}(\vec{N})$ commute. These equations are uniquely solved by 
\begin{equation}\label{supermom}
  \hat{\mathcal{D}}(\vec{N}) = \sum_{I=2}^{\deg P} \int_\Sigma dy\, N^\beta(y) \left[\partial_\beta \hat P^{\alpha_1 \dots \alpha_I} \hat \pi_{\alpha_1 \dots \alpha_I} + I\, \partial_{\alpha_1}(\hat P^{\alpha_1 \dots \alpha_I}\hat  \pi_{\alpha_2 \dots \alpha_I \beta})\right]\,,
\end{equation}
where an a priori non-zero additive integration constant is forced to be zero by (\ref{repDD}). This is already the desired supermomentum appearing in the dynamics (\ref{classHam}). Note that, in the case of $\deg P=2$, the supermomentum reduces to the standard form $\hat{\mathcal{D}}_\alpha = 2 \hat P^{\beta\gamma} \nabla_\beta\hat \pi_{\alpha\gamma}$ known from general relativity \footnote{In the standard treatments of canonical general relativity one chooses the metric $g_{\alpha\beta}$ and its conjugate momentum $\pi^{\alpha\beta}_\textrm{standard}$ as the phase space variables. It can be checked that changing the configuration variables from the metric to its inverse results in the canonically conjugate momentum $\hat\pi_{\alpha\beta}=-g_{\alpha\gamma}g_{\beta\delta}\pi^{\gamma\delta}_\textrm{standard}$,  which explains the different sign in the supermomentum.}.

\subsection{Construction of the non-local superhamiltonian}\label{sec_nonlocal}
Now that the supermomentum, and thus the right hand side of the bracket (\ref{repHH}) is known, we can start to determine the superhamiltonian by extracting the information contained in this and the other remaining bracket (\ref{repDH}). The latter simply tells us that the superhamiltonian must be a scalar density of weight one. For on the one hand, we concluded from (\ref{PonD}) and (\ref{Ponpi}) that $\{\cdot, \hat{\mathcal{D}}(\vec{N})\}$ acts like a Lie derivative on any functional of the geometric phase space variables, and thus on the superhamiltonian in particular. But on the other hand, letting $B(z) = \delta_y(z)$ in the bracket (\ref{repDH}) we obtain
\begin{equation}
  \{\hat{\mathcal{H}}(y), \hat{\mathcal{D}}(\vec{N})\} = \partial_\alpha(\hat{\mathcal{H}}(y) N^\alpha(y))
\end{equation}
after an integration by parts on the right hand side. But this is the Lie derivative of $\hat{\mathcal{H}}$ only if the latter is a scalar density of weight one, as claimed, and this is all that follows from this second bracket.

Again we approach the solution of the Poisson relations starting from one of the necessary relations (\ref{application}). From the left equation we indeed find that
\begin{equation}\label{Hlocaltest}
  \frac{\delta \hat{\mathcal{H}}(N)}{\delta \hat{\pi}_{\alpha_1 \dots \alpha_I}(z)} 
= N(z)\Big[\dots\Big] + \partial_\beta N(z)\Big[(I\min\deg P) \hat P^{\beta\alpha_1 \dots \alpha_I}(z) + (\deg  P\min1)\, I\, \hat P^{(\alpha_2\dots\alpha_I}(z)\hat P^{\alpha_1)\beta}(z)\Big]\,,
\end{equation}
where the expression in $[\dots]$ contains the configuration variables $P^A$, all frame covectors, and their first derivatives. The second term, in contrast, only contains the configuration variables, and this will become important shortly. 
If the second bracket vanished in general (which, however, only is the case for $\deg P = 2$), the superhamiltonian would be a function, rather than a functional, in the momenta $\hat \pi_A$ according to (\ref{Hlocaltest}). However, the simple form of the $\partial_\beta N$ term allows to directly determine the non-local contribution of the momenta to the superhamiltonian. In fact, it is straightforward to check that one can decompose the superhamiltonian as 
\begin{equation}\label{Hdecomposition}
 \hat{\mathcal{H}}(y)[\hat P,\hat \pi] = \hat{\mathcal{H}}_\textrm{local}(y)[\hat P](\hat \pi) + \hat{\mathcal{H}}_\textrm{non-local}(y)(\hat P,\partial\hat \pi)\,,
\end{equation}
namely into a local part $ \hat{\mathcal{H}}_{\textrm{local}}(y)$, which is indeed a functional of $\hat P$ but only a function of $\hat\pi$, and the explicit non-local part  
\begin{equation}\label{Hnonlocal}
  \hat{\mathcal{H}}_{\textrm{non-local}}(y)[\hat P, \hat\pi] = \sum_{I=2}^{\deg P}\big[(\deg P\min I)\partial_{\beta}(\hat P^{\beta\alpha_1\dots\alpha_I}\hat \pi_{\alpha_1\dots\alpha_I}) - (\deg P\min1)\,I\,\partial_{\beta}(\hat P^{\alpha_2\dots\alpha_I}\hat P^{\alpha_1\beta} \hat \pi_{\alpha_1\dots\alpha_I})\big](y)\,,
\end{equation}
which is thus a completely known functional of $\hat P$ and $\hat\pi$ that generates the non-local second term in (\ref{Hlocaltest}). Note that the non-local part $\hat{\mathcal{H}}_{\textrm{non-local}}(y)$ of the superhamiltonian is the divergence of a vector density of weight one and thus a scalar density of the same weight. Hence, the decomposition (\ref{Hdecomposition}) turns the superhamiltonian into the sum of two tensor densities of weight one \cite{GeomKucharI}. This means that we reduced the problem of finding the superhamiltonian as a functional of both phase space variables $\hat P$ and $\hat \pi$ to the much simpler problem, as it will turn out, of determining the local part that is a functional in $\hat P$ but only a function in $\hat \pi$. In particular, this will allow to make a power series ansatz for $\hat{\mathcal{H}}_{\textrm{local}}$ in $\hat\pi$.   

\subsection{Lagrangian reformulation\label{sec_Lagrangian}}
At this point we explicitly know the supermomentum $\hat{\mathcal{D}}$ and the non-local part $\hat{\mathcal{H}}_{\textrm{non-local}}$ of the superhamiltonian (\ref{Hdecomposition}). The still undetermined local part $\hat{\mathcal{H}}_{\textrm{local}}$ of the latter enters the only remaining Poisson bracket (\ref{repHH}) quadratically on its left hand side, 
\begin{equation}\label{quadraticversion}
  \int_\Sigma dz\, \left[\frac{\delta \hat{\mathcal{H}}(x)_{\textrm{local}}}{\delta \hat{P}^A(z)} +\frac{\delta \hat{\mathcal{H}}(x)_{\textrm{non-local}}}{\delta \hat{P}_A(z)}\right]\left[\frac{\delta \hat{\mathcal{H}}(y)_{\textrm{local}}}{\delta \hat{\pi}_A(z)}
  + \frac{\delta \hat{\mathcal{H}}(y)_{\textrm{non-local}}}{\delta \hat{\pi}_A(z)}\right]
  - (x \leftrightarrow y)\,, 
\end{equation} 
where the contributions of the non-local part of the Hamiltonian are explicitly known from taking the functional derivative of (\ref{Hnonlocal}). Here and in the remainder of this paper, repeated multi-indices indicate sums of the form 
\begin{equation}
  C^A D_A = \sum_{I=2}^{\deg P} C^{\alpha_1 \dots \alpha_I} D_{\alpha_1 \dots \alpha_I}\,.
\end{equation}
The quadratic appearance of $\hat{\mathcal{H}}_{\textrm{local}}$ in (\ref{quadraticversion}) seriously complicates a power series ansatz for it in the momenta $\hat \pi$. Remarkably, a Legendre transformation \cite{Kuchar} replacing the momenta $\hat \pi_A$ by Legendre dual variables
\begin{equation}\label{Legendreduals}
   K^A(x) := \frac{\partial \hat{\mathcal{H}}(x)_{\textrm{local}}}{\partial \hat{\pi}_A(x)}\,,
\end{equation}
from which conversely the momenta depend as a function, $\hat{\pi}(x)[P](K)$, allows to turn the equation (\ref{quadraticversion}) that is quadratic in $\hat{\mathcal{H}}_{\textrm{local}}$ into an equation that is linear in the ``Lagrangian''
\begin{equation}\label{Lagrangian}
   L(x)[\hat{P}](K) := \hat{\pi}_A(x)[\hat{P}](K) \,K^A(x) - \hat{\mathcal{H}}(x)_{\textrm{local}}[\hat{P}]\, (\hat{\pi}[\hat{P}](K))\,,
\end{equation}
since then one finds 
\begin{equation}\label{reverse}
  \left.\frac{\delta\hat{\mathcal{H}}(x)_{\textrm{local}}}{\delta \hat{P}^A(y)}\right|_{\hat\pi[\hat{P}](K)} = - \frac{\delta L(x)}{\delta \hat{P}^A(y)}\quad\textrm{ and }\quad \frac{\partial L(x)}{\partial K^A(x)} = \hat{\pi}_A(x)[\hat{P}](K)\,.
\end{equation}
Let us further define the coefficients $Q_{A}{}^{B\beta}$ and $M^{A\zeta}$ by  
\begin{eqnarray}
  \frac{\delta \hat{\mathcal{H}}_{\textrm{non-local}}(x)}{\delta \hat{P}^A(z)}&=& Q_{A}{}^{B\beta}(x)\partial_\beta\hat{\pi}_B(x)\delta_x(z)-Q_{A}{}^{B\beta}(x)\hat{\pi}_B(x)\partial_\beta\delta_x(z)\label{Qdef}\,,\\
\frac{\delta \hat{\mathcal{H}}_{\textrm{non-local}}(y)}{\delta \hat{\pi}_A(z)} &=&  M^{A\zeta}(y)\partial_\zeta\delta_y(z)-\partial_\zeta M^{A\zeta}(y)\delta_y(z)\,,\label{Mdef}
\end{eqnarray}
which yields the expressions
\begin{eqnarray}
  Q_{\alpha_1\dots\alpha_K}{}^{\beta_1\dots\beta_I\, \mu} &=& \delta^K_{I+1} (\deg P \min I) \delta^{\mu \beta_1\dots\beta_I}_{(\alpha_1\dots\alpha_{I+1})}
 - \delta^K_2 I (\deg P\min 1) \hat{P}^{(\beta_2\dots\beta_I} \delta^{\beta_1)\mu}_{(\alpha_1\alpha_2)}\nonumber\\
 &&- \delta^K_{I-1} I (\deg P \min 1) \hat{P}^{\mu(\beta_1}\delta^{\beta_2\dots\beta_I)}_{\alpha_1\dots\alpha_{I-1}}\,,\\
 M^{\alpha_1\dots\alpha_I\, \beta} &=&-(\deg P \min I) \hat{P}^{\beta\alpha_1\dots\alpha_I} + I (\deg P\min 1) \hat{P}^{\beta(\alpha_1} \hat{P}^{\alpha_2\dots\alpha_I)}\,
\end{eqnarray}
depending only on the configuration variables $\hat P$.
Rewriting the Poisson bracket (\ref{quadraticversion}) with the help of the Lagrangian and integrating out the appearing delta distributions, its left hand side takes the form
\begin{eqnarray}
 &-& \frac{\delta L(x)}{\delta \hat{P}^A(y)} K^A(y) + \partial_{y^\zeta}\left[\frac{\delta L(x)}{\delta \hat{P}^A(y)} M^{A\zeta}(y)\right] + M^{A\zeta}(y) Q_{A}{}^{B\beta}(x)\partial_\beta\hat{\pi}_B(x)\partial_\zeta \delta_y(x)\nonumber\\
 &-& K^A(y)Q_A{}^{B\beta}(x)\hat\pi_B(x)\partial_\beta\delta_x(y)+Q_A{}^{B\beta}(x)\hat\pi_B(x)M^{A\xi}(y)\partial^2_{\beta\xi}\delta_y(x)\nonumber\\
&+&Q_A{}^{B\beta}(x)\hat\pi_B(x)\partial_\xi M^{A\xi}(y)\partial_{\beta}\delta_x(y)-(x \leftrightarrow y)\,,\nonumber
\end{eqnarray}
while the right hand side becomes $(\deg P -1)$ times
\begin{eqnarray}
  \sum_{I=2}^{\deg P} \left[\hat{P}^{\beta\alpha}\partial_\beta\hat{P}^{\alpha_1\dots\alpha_I}\hat{\pi}_{\alpha_1\dots\alpha_I} + I \hat{P}^{\beta\alpha}\partial_{\alpha_1}\hat{P}^{\alpha_1\dots\alpha_I}\hat{\pi}_{\alpha_2\dots\alpha_I} + I \hat{P}^{\beta\alpha}\hat{P}^{\alpha_1\dots\alpha_I}\partial_{\alpha_1}\hat{\pi}_{\alpha_2\dots\alpha_I\beta}\right](y)\partial_\alpha \delta_x(y) - (x \leftrightarrow y)\,,\nonumber
\end{eqnarray}
where $\hat \pi$ is given by the second of the equations (\ref{reverse}).
A key observation is now that the dependence of the terms in square brackets on the right hand side may be changed from $y$ to $x$ while the dependence of the delta distribution multiplying it remains unchanged; due to the exchange term $(x \leftrightarrow y)$, the resulting distributions are the same. The same remark applies to changing the dependence of $Q_A{}^{B\beta}$ and $\partial \hat{\pi}_B$ from $x$ to $y$ in the third term on the left hand side. We may thus collect the derivative terms $\partial \pi$ from both sides into an expression of the form 
\begin{equation}\label{partialpiterms}
  T^{A\mu\nu}(x)(\hat P) \partial_\mu\delta_x(y) \partial_\nu\hat \pi_A(x) - (x \leftrightarrow y)
\end{equation}
on the left hand side of the original Poisson bracket relation. 
Crucially, one finds that $T^{A\mu\nu} = T^{A\nu\mu}$ by inspecting the explicit expression for the above coefficients. It is only due to this fact that (\ref{partialpiterms}) is equal to
\begin{equation}
  \left[T^{A\mu\nu}(x)(\hat P) \partial_\mu\partial_\nu \delta_x(y) - \partial_\mu T^{A\mu\nu}(x)(\hat P) \partial_\nu \delta_x(y)\right] \hat\pi_A(x) - (x \leftrightarrow y)
\end{equation}
as a distribution in two variables. Thus all $\partial\pi$ terms can be made into local expressions in the $K$ by virtue of the second relation in (\ref{reverse}).  

Combining all terms of the original Poisson bracket (\ref{repHH}) in this fashion, one obtains its entirely equivalent formulation as a homogeneous linear functional differential equation in $L=L[\hat P](K)$,  
\begin{eqnarray}\label{homolin}
   0 &=& - \frac{\delta L(x)}{\delta \hat{P}^A(y)} K^A(y) + \partial_{y^\zeta}\left[\frac{\delta L(x)}{\delta \hat{P}^A(y)} M^{A\zeta}(y)\right] - \frac{\partial L(x)}{\partial K^A(x)}K^B(x)Q_{B}{}^{A\beta}(x)\partial_\beta\delta_x(y)\nonumber\\
&&+\frac{\partial L(x)}{\partial K^A(x)}\left[ U^{A\mu\nu}(x)\partial^2_{\mu\nu} \delta_x(y) + S^{A\mu}(x) \partial_\mu \delta_x(y) \right] - (x \leftrightarrow y)\,, 
\end{eqnarray}
where the coefficients $U^{A\mu\nu}$ contain the configuration variables,
\begin{equation}\label{Rcoeff}
 U^{\alpha_1\dots\alpha_I\mu\nu}=-I(\deg P\min1)\hat P^{(\mu|(\alpha_1}\hat P^{\alpha_2\dots\alpha_I)|\nu)}\,,
\end{equation}
whereas the coefficients $S^{A\mu}$ also contain their first partial derivatives,
\begin{eqnarray}\label{Scoeff}
S^{\alpha_1\dots\alpha_I\mu}&=&-(\deg P -1)\hat P^{\beta\mu}\partial_{\beta}\hat P^{\alpha_1\dots\alpha_I}+I(\deg P-I)(\deg P-1)\hat P^{\mu(\alpha_1\dots}\partial_\beta  \hat P^{\alpha_I)\beta}\nonumber\\
&&+2I (\deg P -1)\hat P^{(\mu|(\alpha_1\dots}\partial_\beta \hat P^{\alpha_I)|\beta)}+I(\deg P-1)(\deg P-2)\hat P^{\mu\beta(\alpha_1}\partial_\beta\hat P^{\alpha_2\dots\alpha_I)}\nonumber\\
&&-I(I-1)(\deg P -1)^2 \hat P^{\mu(\alpha_1} \hat P^{\alpha_2\dots\alpha_{I-1}}\partial_\beta \hat P^{\alpha_I)\beta},
\end{eqnarray}
where in the case $\deg P=2$ the last term is to be read as $-2 \hat P^{\mu(\alpha_1}\partial_\beta\hat P^{\alpha_2)\beta}$. 
Once one has solved (\ref{homolin}) for the Lagrangian $L[P](K)$, one can recover the momenta
\begin{equation}
  \hat \pi_A(x) = \frac{\partial L(x)[\hat P](K)}{\partial K^A(x)},
\end{equation}
conversely expressing $K^A(x) = K^A(x)[\hat P](\hat \pi)$, and one also finds the local part of the superhamiltonian as
\begin{equation} 
  \hat{\mathcal{H}}(x)_{\textrm{local}}[\hat{P}](\hat{\pi}) =\hat \pi_A(x) K^A(x)[P](\hat\pi) - L(x)[\hat P](K^A(x)[\hat P](\hat \pi))\,. 
\end{equation}
This then of course amounts to the full determination of the gravitational dynamics, since the supermomentum and non-local part of the superhamiltonian are already known from previous sections. But the difficulty of solving (\ref{homolin}) consists in this being a distributional functional differential equation for $L$. 

\subsection{Reduction to differential equations\label{sec_reduction}}
In this section we will reduce the equation (\ref{homolin}) to a countable set of linear partial differential equations for the functional $L$ that determines the still missing local part of the superhamiltonian. 
This reduction takes place in two steps:
\begin{center}
distributional functional  differential equation\\
$\downarrow$\\
distributional differential equations\\
$\downarrow$ \\
differential equations
\end{center} 

The first step to achieve this exploits the linear homogeneous structure of the equation (\ref{homolin}) by making a power series ansatz
\begin{equation}\label{Kseries}
  L(x)[\hat{P}](K) = \sum_{i=0}^\infty C(x)[\hat{P}]_{A_1\dots A_i} K^{A_1}(x) \dots K^{A_i}(x) 
\end{equation}
with coefficients that are so far undetermined functionals of, and this is the essential point, only the configuration variables. A power series expansion is justified since we took care in constructing $L$ as a mere function of $K$, while it remains a functional of $\hat P$. Since the velocities $K^A$ are defined as the partial derivatives of the weight-one scalar density $H_{local}$ with respect to the tensor densities $\hat\pi_A$ of the same weight, the velocities themselves are tensors. The coefficient functionals $C[P]_{A_1 \dots A_N}$ are thus tensor densities of weight one just as the Lagrangian $L[P](K)$. 

Insertion of (\ref{Kseries}) into (\ref{homolin}) replaces the latter, a distributional differential equation for $L[P](K)$, by a countable set of such equations for the coefficient functionals $C[P]_{A_1 \dots A_N}$; one equation for each order $N$ in $K$.  
Extracting the $N$-th order equation by application of the functional derivative operator
\begin{equation}
  \frac{\delta^N}{\delta K^{B_1}(x_1) \dots \delta K^{B_N}(x_N)}
\end{equation}
to the equation (\ref{homolin}) and evaluating the result at $K=0$, we will now see that one obtains at $N$-th order a distributional equation in $N+2$ variables $x,y,x_1,\dots,x_N$. Indeed, the zeroth order contribution is 
\begin{equation}\label{fde0}
  0 = \partial_{y^\zeta}\left[M^{A\zeta}(y)  \frac{\delta C(x)}{\delta P^A(y)}\right] 
 + C(x)_A \left[U^{A\mu\nu}(x)\partial_\mu\partial_\nu \delta_x(y) + S^{A\mu}(x) \partial_\mu \delta_x(y) \right] - (x \leftrightarrow y)\,,
\end{equation}
while the contribution at order $N\geq 1$ is
\begin{eqnarray}
 0 &=& \quad \Big\{(N+1)! G(x)_{A B_1 \dots B_N}\left(U^{A\mu\nu}(x)\partial_\mu\partial_\nu \delta_x(y) + S^{A\mu}(x) \partial_\mu \delta_x(y)\right)\nonumber\\
  & & \,\,\quad + N!\, \partial_{y^\zeta}\left[M^{A\zeta}(y)  \frac{\delta C(x)_{B_1 \dots B_N}}{\delta P^A(y)}\right]-NN!Q_{(B_1}{}^{M\beta}(x)C(x)_{B_2\dots B_N)M}\partial_\beta\delta_x(y)\Big\}\,  \delta_x(x_1) \dots \delta_x(x_N)\nonumber\\
  & & \,\,\, - (N-1)! \, \sum_{j=1}^N  \frac{\delta C(x)_{B_1 \dots \widetilde{B_j} \dots B_N}}{\delta P^{B_j}(y)} \delta_y(x_j) \, \delta_x(x_1) \dots \widetilde{\delta_x(x_j)} \dots \delta_x(x_N)\label{fdeN}\,,
\end{eqnarray}
where $\sim$ instructs to omit a term. As usual, these distributional equations are to be understood by first applying them to test functions $f(x, y, x_1, \dots, x_N)$ and then integrating over all variables. In order to convert the thus constructed functional differential equations into regular differential equations, we restrict attention to coefficients $C(x)_{A_1 \dots A_i}[\hat{P}]$ that are determined by the value of $\hat{P}$ and all its derivatives at $x$, so that 
\begin{equation}
   C(x)[\hat{P}]_{A_1 \dots A_i} = C_{A_1 \dots A_i}(\hat{P}(x), \partial\hat{P}(x), \partial\partial\hat{P}(x), \dots)\,.
\end{equation}
This allows, in particular, to write 
\begin{equation}\label{weaklylocal}
  \frac{\delta C(x)[\hat{P}]_{B_1 \dots B_i}}{\delta P^A(y)} = \sum_{j=0}^\infty (-1)^j \frac{\partial C(x)_{B_1 \dots B_i}(\hat{P},\partial\hat{P},\dots)}{\partial \partial^j_{\alpha_1\dots\alpha_j} \hat{P}^A(x)} \partial^j_{\alpha_1\dots \alpha_j} \delta_x(y)\,.
\end{equation}
in the functional differential equations (\ref{fde0}) and (\ref{fdeN}). This completes the first step of the reduction process of equation (\ref{homolin}), to a countable set of distributional differential equations. 

The strategy to convert these into regular differential equations now begins with eliminating all $\delta$ distributions, which requires to shovel derivatives over to the test function. For the zeroth order equation (\ref{fde0}) we obtain
\begin{eqnarray}\label{zerothintegrated}
   0 = \int dx \Big\{ & & C_A U^{A\mu\nu}(x)(\partial^2_{2\,\mu\nu} f)(x,x) - C_A(x) S^{A\mu}(x) (\partial_{2\,\mu}f)(x,x)\\
 & & - \sum_{j=0}^\infty \sum^{j}_{s=0} \binom{j}{s} \frac{\partial C(x)}{\partial\partial^j_{\alpha_1 \dots \alpha_j}\hat{P}^A(x)}(\partial^{s+1}_{2\,\zeta(\alpha_1\dots\alpha_{s}} f)(x,x) (\partial^{j-s}_{\alpha_{s+1}\dots\alpha_j)} M^{A\zeta})(x)\Big\} - (\partial_2\rightarrow \partial_1)\nonumber\,
\end{eqnarray}  
for any test function $f(x,y)$ with compact support. Unfortunately, one cannot directly read off from this equation that the coefficient functions of the various derivatives of $f$ all vanish. This is because the derivatives $\partial_1 f$ and $\partial_2f$ of the test function are evaluated at $(x,x)$ rather than $(x,y)$, and thus are not independent of each other. Indeed, we have $\partial_\mu f(x,x) = (\partial_{1\,\mu} f)(x,x) + (\partial_{2\,\mu} f)(x,x)$, so that 
\begin{eqnarray}
 & & \int dx \big\{A(x) f(x,x) + B^\mu(x) (\partial_{1\,\mu}f)(x,x) + C^\mu(x) (\partial_{2\,\mu} f)(x,x)\big\}\nonumber\\
 &=& \int dx \big\{[A(x)-\partial_\mu C^\mu(x)]f(x,x) + [B^\mu(x)-C^\mu(x)](\partial_{1\,\mu} f)(x,x)\big\}\,.
\end{eqnarray}    
In particular, the vanishing of the first integral for any arbitrary test function $f$ only implies that $A-\partial_\mu C^\mu = 0$ and $B^\mu-C^\mu = 0$, but not that the coefficient functions $A$, $B^\mu$ and $C^\mu$ would vanish individually. This applies similarly if higher order derivatives are involved, since with
\begin{equation}\label{convert2to1}
  (\partial_{2\, \alpha_1\dots\alpha_n}^n f)(x,x) = \sum_{t=0}^{n} \tbinom{n}{t} (-1)^t (\partial^{n-t}_{(\alpha_1\dots\alpha_t}\partial_{1\, \alpha_{t+1}\dots\alpha_n)}^t f)(x,x) 
\end{equation}
we can always express derivatives acting on the second entry of $f$ by those acting on the first entry and total derivatives, and then read off the independent equations. Using (\ref{convert2to1}) and re-ordering multiple sums, the zeroth order equation (\ref{zerothintegrated}) can be brought to the form
\begin{equation}
  0 = \int dx\, \Big\{ f(x,x) A(x) + \sum_{w=1}^\infty (\partial^w_{1\,\beta_1\dots\beta_w}f)(x,x) B^{\beta_1 \dots \beta_w}(x)\Big\}
\end{equation}
where the vanishing of the coefficient $A$ amounts to the differential equation
\begin{equation}
\tbinom{N=0}{w=0} \quad
  0 = \partial^2_{\mu\nu}(C_A U^{A \mu\nu}) + \partial_\mu(C_A S^{A\mu}) - \sum_{j=0}^\infty\sum_{s=0}^{j} (-1)^s \binom{j}{s} \partial^{s+1}_{\zeta\alpha_1\dots\alpha_{s}}\left[\frac{\partial C}{\partial\partial^j_{\alpha_1\dots\alpha_j}\hat{P}^A} \partial^{j-s}_{\alpha_{s+1}\dots\alpha_j} M^{A\zeta}\right],\nonumber
\end{equation}
the vanishing of the coefficient $B^{\beta_1}$ to the differential equation
\begin{eqnarray}
 \tbinom{N=0}{w=1}\quad  0 &=& 2 \partial_\mu(C_A R^{A\beta\mu}) + 2 C_A S^{A\beta_1} + \sum_{j=0}^\infty \partial^j_{\gamma_1\dots\gamma_j}M^{A\beta} \frac{\partial C}{\partial\partial^j_{\gamma_1\dots\gamma_j} \hat{P}^A} \nonumber\\
   & & + \sum_{j=0}^\infty \sum_{s=0}^{j} (-1)^s \tbinom{j}{s} (s+1) \,\partial^{s}_{\alpha_1\dots\alpha_{s}}\left[\partial^{j-s}_{\gamma_1\dots\gamma_{j-s}} M^{A(\beta|} \frac{\partial C}{\partial\partial^j_{|\alpha_1\dots\alpha_s)\gamma_1\dots\gamma_{j-s}}\hat{P}^A}\right]\,,\qquad\qquad\qquad\qquad\qquad\nonumber
\end{eqnarray}
and the vanishing of all further coefficients $B^{\beta_1\beta_2\dots}$ to the differential equations
\begin{eqnarray}
 \tbinom{N=0}{w\geq 2}\quad  0 &=& \sum_{j=w-1}^\infty \tbinom{j}{w-1} \partial^{j+1-w}_{\alpha_w\dots\alpha_j}M^{A(\beta_1|} \frac{\partial C}{\partial\partial^j_{|\beta_2 \dots \beta_w) \alpha_w \dots \alpha_j}\hat{P}^A} \nonumber\\
  &+&  \sum_{j=w-1}^\infty \sum_{s=w-1}^{j} (-1)^s \tbinom{j}{s}\tbinom{s+1}{w} \partial^{s+1-w}_{\alpha_1\dots\alpha_{s+1-w}}\left[\partial^{j-s}_{\gamma_1\dots\gamma_{j-s}}M^{A(\alpha_{s+1-w}|} \frac{\partial C}{\partial\partial^j_{|\beta_1 \dots \beta_w \alpha_1 \dots \alpha_{s-w})\gamma_1\dots{\gamma_{j-s}}} \hat{P}^A}\right]\,.\nonumber\qquad
\end{eqnarray}
This countable set of partial differential equations for the coefficients $C$ and $C_A$ is equivalent to the information contained in the one functional differential equations (\ref{fde0}) arising at order $N=0$ in $K$. 

Similarly, one obtains for each order $N\geq 1$ from equation (\ref{fdeN}) first the distributional differential equation
\begin{eqnarray}\label{fdeNinter}
  0 &=& \int dx \, \Big\{
      (N+1)! C_{AB_1\dots B_N} \left(U^{ A\gamma\delta}\partial^2_{2\,\gamma\delta} f - S^{ A\gamma}\partial_{2\,\gamma} f \right)+NN! Q_{B_1}{}^{M\beta}C_{B_2\dots B_N)M}\partial_{2\,\beta}\nonumber\\\
   & & \qquad\quad - N! \sum_{s=0}^\infty \sum_{j=s}^\infty \binom{j}{s} \frac{\partial C_{B_1\dots B_N}}{\partial\partial^j_{\alpha_1\dots \alpha_j}\hat{P}^A}\left(\partial^{s+1}_{2\,\zeta(\alpha_1\dots\alpha_{s-1}} f \partial^{j-s}_{\alpha_{s+1} \dots \alpha_j)} M^{A \zeta} \right) \nonumber\\
  & & \qquad\quad -(N-1)! \sum_{s=1}^\infty\sum_{j=s}^\infty\sum_{i=1}^{N-1}  \frac{\partial C_{B_1 \dots \widetilde{B_i} \dots B_N}}{\partial\partial^j_{\alpha_1 \dots \alpha_j}\hat{P}^{B_i}} \partial^s_{2\, (\alpha_1\dots \alpha_s}\partial^{j-s}_{(i+2)\,\alpha_{s+1}\dots\alpha_j)} f\nonumber\\
  & & \qquad\quad +(N-1)! \sum_{t=1}^\infty\sum_{k=0}^\infty \sum_{j=k+t}^{\infty}(-1)^j  \frac{j!}{t!k!(j\min t\min k)!} \partial^{j-t-k}_{\alpha_1\dots\alpha_{j-t-k}} \frac{\partial C_{B_1 \dots B_{N-1}}}{\partial\partial^j_{\alpha_1\dots\alpha_j}\hat{P}^{B_N}}\times\nonumber\\
  & & \qquad\qquad\qquad\qquad\qquad\qquad\qquad\times\partial^{t}_{2\,\alpha_{j\min t\min k}\dots\alpha_{j\min k+1}}\partial^k_{(3,\dots, N+1)\alpha_{j\min k +2}\dots\alpha_j} f \Big\} - \Big\{\partial_2\to \partial_1\Big\} \,,\label{higherintegrated}
\end{eqnarray}
where $\partial_{(3,\dots, N+1)}f$ denotes a derivative acting only on entries three to $N+1$ of the test function. The last multiple sum arises from an elimination of the partial derivatives acting on entry number $N+2$ of the test function by way of the identity 
\begin{equation}
  \partial^j_{(2,N+2) \, \alpha_1\dots \alpha_j} f = \sum^j_{s=0} \tbinom{j}{s} \partial^s_{(\alpha_1 \dots \alpha_s} (-1)^{j-s} \partial^{j-s}_{(1,3,\dots,N+1) \, \alpha_{s+1} \dots \alpha_j} f\,,
\end{equation}
which renders the distributional differential equations (\ref{higherintegrated}) for each $N$ free of derivatives $\partial_{N+2} f$ and thus removes ambiguities due to surface terms, so that one can now write (\ref{fdeNinter}) in the form
\begin{equation}
  0 = \int dx\, \sum_{s=1}^\infty \sum_{j=0}^\infty \sum_{\text{Part}_m(j)} {}^{(s;j)}B^{\beta_1 \dots \beta_{s+j}}_{B_1 \dots B_N} (\partial^s_2 \partial^{m_3}_{3} \dots \partial^{m_{N+1}}_{N+1})_{(\beta_1 \dots \beta_{s+j})}f \,-\, (\partial_2\to\partial_1)\,,
\end{equation}
where the third sum is meant as the sum over partitions $j=m_3+\dots m_{N+1}$. Employing various multinomial distributions of higher derivatives and reordering of sums one obtains the following equations for $N\geq 1$. At level $j=0$ one obtains from the vanishing of the coefficient ${}^{(1;0)}B$ the equation
\begin{eqnarray}
 \tbinom{N\geq 1}{s=1; j=0 + \dots + 0} \qquad 0 &=& (N+1)! C_{AB_1 \dots B_N} S^{A\beta} - NN!Q_{(B_1}{}^{M\beta}C_{B_2\dots B_N)M}\nonumber\\
  & &+ N! \sum_{j=0}^\infty \frac{\partial C_{B_1\dots B_N}}{\partial\partial^j_{\alpha_1\dots\alpha_j}\hat{P}^A} \partial^j_{\alpha_1 \dots \alpha_j} M^{A\beta} + (N-1)! \sum_{i=1}^{N-1} \frac{\partial C_{B_1 \dots \widetilde{B_i} \dots B_N}}{\partial\partial_\beta \hat{P}^{B_i}} \nonumber\\
  & &- (N-1)! \sum_{j=1}^\infty (-1)^j j \partial^{j-1}_{\alpha_2 \dots \alpha_j} \frac{\partial C_{B_1 \dots B_{N-1}}}{\partial\partial^j_{\beta \alpha_2 \dots \alpha_j} \hat{P}^{B_N}}\,,\qquad\qquad\qquad\qquad\qquad\qquad\qquad\qquad\nonumber
\end{eqnarray}
from the vanishing of the coefficient ${}^{(2;0)}B$ the equation
\begin{eqnarray}
 \tbinom{N\geq 1}{s=2; j=0 + \dots + 0} \qquad 0 &=& (N+1)! C_{AB_1 \dots B_N} U^{A\beta_1\beta_2 } - N! \sum_{j=1}^\infty j \frac{\partial C_{B_1\dots B_N}}{\partial\partial^j_{(\beta_1|\alpha_2\dots\alpha_j}\hat{P}^A} \partial^{j-1}_{\alpha_2 \dots \alpha_j} M^{A|\beta_2)}\nonumber\\
  & & -(N-1)! \sum_{i=1}^{N-1} \frac{\partial C_{B_1 \dots \widetilde{B_i} \dots B_N}}{\partial\partial^2_{\beta_1\beta_2} \hat{P}^{B_i}} + (N-1)! \sum_{j=2}^\infty (-1)^j \tbinom{j}{2} \partial^{j-2}_{\alpha_3 \dots \alpha_j} \frac{\partial C_{B_1 \dots B_{N-1}}}{\partial\partial^j_{\beta_1\beta_2 \alpha_3 \dots \alpha_j} \hat{P}^{B_N}}\,,\qquad\qquad\nonumber
\end{eqnarray}
and from the vanishing of the coefficients ${}^{(s\geq3;0)}B$ the equations
\begin{eqnarray}
 \tbinom{N\geq 1}{s\geq 3; j=0 + \dots + 0} \qquad 0 &=&  N! \sum_{j=s-1}^\infty\tbinom{j}{s-1} \frac{\partial C_{B_1\dots B_N}}{\partial\partial^j_{(\beta_1 \dots \beta_{s-1}| \alpha_s\dots\alpha_j}\hat{P}^A} \partial^{j-s+1}_{\alpha_s \dots \alpha_j} M^{A|\beta_s)}\nonumber\\
  & & + (N-1)! \sum_{i=1}^{N-1} \frac{\partial C_{B_1 \dots \widetilde{B_i} \dots B_N}}{\partial\partial^s_{\beta_1\dots\beta_s} \hat{P}^{B_i}} - (N-1)! \sum_{j=s}^\infty (-1)^j \tbinom{j}{s} \partial^{j-s}_{\alpha_{s+1} \dots \alpha_j} \frac{\partial C_{B_1 \dots B_{N-1}}}{\partial\partial^j_{\beta_1 \dots \beta_s \alpha_{s+1}\dots \alpha_j} \hat{P}^{B_N}}\,.\qquad\qquad\nonumber
\end{eqnarray}
At level $j>0$ there are two more types of coefficients that lead to equations. The first type is ${}^{(s\geq1;j=m_{a+2})}B$, where the $a$-th member of the partition $m_{a+2}=j$, and their vanishing leads to the equations
\begin{eqnarray}
 \tbinom{N\geq 1}{s\geq 1; j=0 + \dots + j + \dots + 0} \qquad 0 &=& (N-1)! \tbinom{s+j}{s}\frac{\partial C_{B_1 \dots \widetilde{B_a} \dots B_N}}{\partial\partial^{s+j}_{\beta_1\dots\beta_{s+j}} \hat{P}^{B_a}}\nonumber\\
& & - (N-1)! \sum_{q=s+j}^\infty (-1)^q \frac{q!}{s!j!(q\min j\min s)!} \partial^{q-j-s}_{\alpha_{s+j+1} \dots \alpha_q} \frac{\partial C_{B_1 \dots B_{N-1}}}{\partial\partial^q_{\beta_1 \dots \beta_{s+j} \alpha_{s+j+1}\dots \alpha_q} \hat{P}^{B_N}}\,,\qquad\qquad\nonumber
\end{eqnarray}
and the second type ${}^{(s\geq1;j=\text{Part}_m(j))}B$ covers all remaining partitions of $j\geq 2$, which have at least two non-vanishing members, and their vanishing leads to the equations
\begin{eqnarray}
 \tbinom{N\geq 1}{s\geq 1; j= m_3 + \dots + m_{N+1}} \qquad 0 &=&
- \frac{(N-1)!}{m_3! \dots m_{N+1}!} \sum_{q=s+j}^\infty (-1)^q \frac{q!}{s!(q\min j\min s)!} \partial^{q-j-s}_{\alpha_{s+j+1} \dots \alpha_q} \frac{\partial C_{B_1 \dots B_{N-1}}}{\partial\partial^q_{\beta_1 \dots \beta_{s+j} \alpha_{s+j+1}\dots \alpha_q} \hat{P}^{B_N}}\,.\qquad\qquad\nonumber
\end{eqnarray}
Fortunately, these equations encoding the first Poisson bracket relation \ref{repHH} considerably simplify upon further inspection, as we will show in the the following section, where they will also be supplemented by equations equivalent to the remaining second Poisson bracket relation (\ref{repDH}). 
   
\subsection{Construction of the local superhamiltonian\label{sec_construction}}
The differential equations for the coefficients $C_{B_1 B_2 \dots}$ imply that the latter only depend on at most second order derivatives of the $P^A$. For one first observes that insertion of equations $(N\geq 1, s\geq 2, m_3+\dots+m_{N+1}\geq 2)$ into the equations $(N\geq 1, s\geq 1, j=0+\dots + j + \dots 0\geq 2)$ yield
\begin{equation}\label{nohigherderivs}
  \frac{\partial C_{B_1\dots \widetilde{B_a} \dots B_N}}{\partial\partial^{s+j}_{\gamma_1\dots\gamma_{s+j}}\hat{P}^{B_a}} = 0\,,
\end{equation} 
first apparently restricted to $N\geq 1$, but then insertion of this result into the difference of equations $(N\geq 1, s=2, j=1)$ and $(N\geq 1, s=3, j=1)$ shows that (\ref{nohigherderivs}) holds in fact for all $N\geq 0$. The only other conclusion one may draw from the last two sets of equations of the previous section is that for $a=1,\dots, N$ we have the symmetry condition
\begin{equation}\label{mainone}
  \frac{\partial C_{B_1 \dots \widetilde{B_a} \dots B_N}}{\partial \partial^2_{\gamma_1\gamma_2}\hat{P}^{B_a}} = \frac{\partial C_{B_1 \dots \dots B_{N-1}}}{\partial \partial^2_{\gamma_1\gamma_2}\hat{P}^{B_N}} \qquad \textrm{for all } N\geq 1\,.
\end{equation}
Insertion of these strong results into the remaining three sets of equations for $N\geq 1$ collapses the latter to two equations coupling coefficients of orders $N+1$, $N$ and $N-1$,
\begin{eqnarray}\label{maintwo}
  0 = & & (N+1)!\, C_{AB_1\dots B_N} U^{A\alpha\beta } - N!\, \frac{\partial C_{B_1\dots B_N}}{\partial\partial_{(\beta|} \hat{P}^A} M^{A|\alpha)} - 2 N!\, \frac{\partial C_{B_1\dots B_N}}{\partial\partial^2_{(\beta|\gamma}\hat{P}^A} \partial_\gamma M^{A|\alpha)}  \nonumber\\
  & & -(N-2) (N-1)! \frac{\partial C_{B_1 \dots B_{N-1}}}{\partial\partial^2_{\alpha\beta} \hat{P}^{B_N}}\,,  
\end{eqnarray}
and
\begin{eqnarray}\label{mainthree}
  0 = & & (N+1)! C_{A B_1 \dots B_N} S^{A\alpha } 
           + (N-1)! \sum_{a=1}^N \frac{\partial C_{B_1 \dots \widetilde{B_a} \dots B_N}}{\partial\partial_{\alpha} \hat{P}^{B_a}} 
           - 2(N-1)! \partial_\gamma  \frac{\partial C_{B_1 \dots B_{N-1}}}{\partial\partial^2_{\alpha\gamma} \hat{P}^{B_N}}\nonumber\\
   & & + N! \frac{C_{B_1\dots B_N}}{\partial \hat{P}^A} M^{A \alpha}        
       + N! \frac{\partial C_{B_1\dots B_N}}{\partial\partial_{\gamma} \hat{P}^A} \partial_\gamma M^{A\alpha} 
       + N! \frac{\partial C_{B_1 \dots B_N}}{\partial\partial^2_{\gamma\delta} \hat{P}^{A}} \partial^2_{\gamma\delta} M^{A\alpha}\,\nonumber\\
   & &   - NN!Q_{(B_1}{}^{M\alpha}C_{B_2\dots B_N)M}.
\end{eqnarray}
as well as a further symmetry condition
\begin{equation}\label{mainfour}
  0 = \frac{\partial C_{B_1 \dots B_N}}{\partial\partial^2_{(\alpha\beta|} \hat{P}^A} M^{A|\gamma)}\qquad \textrm{for all } N\geq 0\,,
\end{equation}
where the $N=0$ case is provided by the equation $(N=0, w=3)$. The only other independent equation is the one for $(N=0, w=1)$, coupling $C$ to $C_A$,
\begin{eqnarray}\label{mainfive}
  0 = & & 2 \partial_\mu(C_A U^{A\beta\mu}) +2 C_A S^{A\beta} - 2 \partial_\mu\left(\frac{\partial C}{\partial\partial_{(\mu|}\hat{P}^A} M^{A|\beta)}\right) - 4 \partial_\mu\left(\frac{\partial C}{\partial\partial^2_{(\mu|\nu} \hat{P}^A} \partial_{\nu}M^{A|\beta)}\right)\nonumber \\
   & & + 2 M^{A\beta}\frac{\partial C}{\partial \hat{P}^A} + 2\partial_\mu M^{A\beta} \frac{\partial C}{\partial\partial_{\mu}\hat{P}^A} + 2 \partial^2_{\mu\nu} M^{A\beta} \frac{\partial C}{\partial\partial^2_{\mu\nu}\hat{P}^A}\,,
\end{eqnarray} 
since the equation $(N=0,w=0)$ is simply the divergence of this, and all equations $(N=0,w\geq 4)$ are identically satisfied. Thus only the five sets of equations (\ref{mainone}),  (\ref{maintwo}), (\ref{mainthree}), (\ref{mainfour}) and (\ref{mainfive}) must be solved for the coefficients $C_{A_1\dots A_N}(\hat P,\partial \hat P,\partial^2\hat P)$.

But in addition to these equations, the weight-one densities $C_{B_1\dots B_N}(\hat P,\partial \hat P,\partial^2\hat P)$ must also satisfy three additional conditions \cite{Rund} imposed by their transformation properties under changes of coordinates on the hypersurface $\Sigma$ (equivalently, these follow from the Poisson bracket of the supermomentum and the superhamiltonian). Under an arbitrary change of coordiantes $\bar x^\alpha=\bar x^\alpha(x)$, the fields $\hat P^{\alpha_1\dots\alpha_I}$ transform as
\begin{equation}\label{transformP}
\bar P^{\beta_1\dots\beta_I}=\hat P^{\alpha_1\dots\alpha_I}(A^{-1})_{\alpha_1}^{\beta_1} \dots (A^{-1})_{\alpha_I}^{\beta_I},
\end{equation}
where $(A^{-1})_{\alpha}^{\beta}=\partial x^\beta/\partial \bar x^\alpha$ is the inverse of the Jacobian $A_{\beta}^{\alpha}=\partial \bar x^\alpha/\partial x^\beta$ of the transformation. Since the coefficients $C_{B_1\dots B_N}[\hat P]$ are all tensor densities of weight one, they transform as
\begin{equation}\label{transformG}
\bar C_{C_1\dots C_N}(\bar P,\bar\partial \bar P,\bar \partial^2\bar P)=\det(A)A_{C_1}^{B_1}\dots A_{C_N}^{B_N}C_{B_1\dots B_N}(\hat P,\partial \hat P,\partial^2\hat P), 
\end{equation}
where $A_{C}^{B}=A_{\gamma_1}^{(\beta_1}\dots A_{\gamma_I}^{\beta_I)}$ denotes the transformation of the capital multi-indices. Taking the derivative of equation (\ref{transformG}) with respect to $(A^{-1})^{\rho}_{\sigma,\mu\nu}=\partial^3 x^{\rho}/(\partial \bar x^{\sigma}\partial \bar x^{\mu}\partial \bar x^{\nu})$, noting that its right hand side is independent of these quantities, we obtain quite generally
\begin{equation}\label{invarone}
0=\sum_{I=2}^{\deg P}~I~\hat P^{\alpha_2\dots\alpha_I (\sigma}\frac{\partial C_{B_1\dots B_N}}{\partial\partial^2_{\mu\nu)} \hat P^{\alpha_2\dots\alpha_I\rho}}.
\end{equation}
This is the first invariance identity for the coefficients $C_{A_1\dots A_N}$ that also follows directly from the constraint algebra. Taking the derivative of equation (\ref{transformG}) with respect to $(A^{-1})^{\rho}_{\mu,\nu}=\partial^2 x^{\rho}/(\partial \bar x^{\mu}\partial \bar x^{\nu})$ and using the first invariance identity (\ref{invarone}) we obtain a second invariance identity:
\begin{equation}\label{invartwo}
0=\sum_{I=2}^{\deg P}\left[I~\hat P^{\alpha_2\dots\alpha_I (\mu}\frac{\partial C_{B_1\dots B_N}}{\partial \partial_{\nu)} \hat P^{\alpha_2\dots\alpha_I\rho}}-\partial_\rho \hat P^{\alpha_1\dots\alpha_I}\frac{\partial C_{B_1\dots B_N}}{\partial\partial^2_{\mu\nu}\hat P^{\alpha_1\dots\alpha_I}}+2I~\partial_{\sigma}\hat P^{\alpha_2\dots\alpha_I(\mu}\frac{\partial C_{B_1\dots B_N}}{\partial\partial^2_{\nu)\sigma}\hat P^{\alpha_2\dots\alpha_I\rho}}\right].
\end{equation}
The last invariance identity is obtained by taking the derivative of (\ref{transformG}) with respect to $(A^{-1})^{\mu}_{\rho}=\partial x^{\rho}/\partial \bar x^{\mu}$ which results in
\begin{eqnarray}\label{invarthree}
&&- \delta^{\rho}_{\mu} C_{B_1\dots B_N}-n_1\delta^{\rho}_{(\beta^{(1)}_1}C_{\beta^{(1)}_2\dots \beta^{(1)}_{n_1})\mu B_2\dots B_N}-\dots -n_NC_{B_1\dots B_{N-1}\mu (\beta^{(N)}_2\dots \beta^{(N)}_{n_N}}\delta^{\rho}_{\beta^{(N)}_1)}\nonumber\\
&&=\sum_{I=2}^{\deg P}\Big [I~\hat P^{\rho\beta_2\dots\beta_I}\frac{\partial C_{B_1\dots B_N}}{\partial \hat P^{\beta_2\dots\beta_I\mu}}+I~\partial_\gamma\hat P^{\rho\beta_2\dots\beta_I}\frac{\partial C_{B_1\dots B_N}}{\partial \partial_\gamma \hat P^{\beta_2\dots\beta_I\mu}} \nonumber\\
&&-\partial_\mu\hat P^{\beta_1\dots\beta_I}\frac{\partial C_{B_1\dots B_N}}{\partial \partial_\rho \hat P^{\beta_1\dots\beta_I}}+I~\partial_{\gamma\delta}\hat P^{\rho\beta_2\dots\beta_I}\frac{\partial C_{B_1\dots B_N}}{\partial \partial_{\gamma\delta} \hat P^{\beta_2\dots\beta_I\mu}}\nonumber\\
&&- 2\partial_{\mu\gamma}\hat P^{\beta_1\dots\beta_I}\frac{\partial C_{B_1\dots B_N}}{\partial \partial_{\rho\gamma} \hat P^{\beta_1\dots\beta_I}}\Big],
\end{eqnarray}
where $n_i$ is the number of small indices contained in the capital index $B_i$ and $B_i=\beta^{(i)}_1\dots \beta^{(i)}_{n_i}$. If we contract the indices ${}^\rho{}_\mu$ we get the simpler indentity
\begin{eqnarray}\label{invarthreecontracted}
-(\dim \Sigma+n) C_{B_1\dots B_N}=\sum_{I=2}^{\deg P}\Big [&I&~\hat P^{\beta_1\dots\beta_I}\frac{\partial C_{B_1\dots B_N}}{\partial \hat P^{\beta_1\dots\beta_I}}+(I-1)~\partial_\gamma\hat P^{\beta_1\dots\beta_I}\frac{\partial C_{B_1\dots B_N}}{\partial \partial_\gamma \hat P^{\beta_1\dots\beta_I}}\nonumber\\
&+&(I-2)~\partial_{\gamma\delta}\hat P^{\beta_1\dots\beta_I}\frac{\partial C_{B_1\dots B_N}}{\partial \partial_{\gamma\delta} \hat P^{\beta_1\dots\beta_I}}\Big],
\end{eqnarray}
with $n$ being the total number of lower case indices contained in all capital indices $B_1$ to $B_N$. Equations (\ref{invarone}),(\ref{invartwo}) and (\ref{invarthree}) together with the equations (\ref{mainone}),  (\ref{maintwo}), (\ref{mainthree}), (\ref{mainfour}) and (\ref{mainfive}) must now completely determine the coefficients $C_{B_1\dots B_N}$. 
These then yield the local part of the superhamiltonian, so that together with the already explicitely known non-local part and  supermomentum, this determines the gravitational dynamics.
The physical problem of finding dynamics for modified dispersion relations is thus reduced to the mere technical problem to solve this set of homogeneous linear partial differential equations.

\subsection{Reduction to first derivative order\label{sec_firstderivative}}
We remark that the linear partial differential equations determining the local part of the super-hamiltonian can in fact be reduced to linear partial differential equations for quantities that depend at most on the $\hat P^A$ and their first partial derivatives. This follows essentially from the observation  that the coefficients $C_{B_1 B_2 \dots}$ depend first of all only polynomially on the second partial derivatives of the $\hat P^A$, and indeed at most to order $\dim M - 1$. Since due to the fact that the coefficients $C_{B_1\dots B_N}$ do not depend on derivatives of $\hat P^A$ higher than the second, we can first extract a further set of equations from (\ref{mainthree}). Writing out the total divergence of the third term, we then conclude that
\begin{equation}\label{symmi}
\frac{\partial C_{B_1\dots B_N}}{\partial \partial^2_{(\rho\sigma}\hat P^C\,\partial\partial^2_{\mu)\nu}\hat P^D}=0,~~~N\geq 0.
\end{equation}
For transparency, we restrict the following technical discussion to the case where $\Sigma$ is a three-dimensional manifold. However, the argument holds in a generalized form in any dimension. Since the coefficients $C_{B_1\dots B_N}$ are tensor denisities it can be checked that for all $N\geq0$
\begin{equation}\label{Lambdas}
\Lambda_{B_1\dots B_N~Q\,\,\,\, R\,\,\,\, S\,\,\,\, T}^{~~~~~~~~~~\alpha\beta~\gamma\delta~\kappa\lambda~\rho\sigma}:=\frac{\partial^4 C_{B_1\dots B_N}}{\partial \partial^2_{\alpha\beta}\hat P^Q\,\partial \partial^2_{\gamma\delta}\hat P^R\,\partial \partial^2_{\kappa\lambda}\hat P^S\,\partial \partial^2_{\rho\sigma}\hat P^T}
\end{equation}
are also components of a tensor density. According to equation (\ref{symmi}) the quantities $\Lambda$ vanish whenever we symmetrise over three adjacent greek indices, which also implies that the $\Lambda$ are totally symmetric under the exchange of the pairs $\alpha\beta$, $\gamma\delta$, $\kappa\lambda$ and $\rho\sigma$. Moreover, the $\Lambda$'s are also totally symmetric under the exchange of $Q,R,S,T$. Let us now investigate all the above components. In three dimensions it is clear that at least three of the eight greek indices in (\ref{Lambdas}) take the same value. Using all the described symmetries we can always arrange for these equal indices to appear right next to each other, which immediately implies that
\begin{equation}
\Lambda_{B_1\dots B_N~Q\,\,\,\, R\,\,\,\, S\,\,\,\, T}^{~~~~~~~~~~\alpha\beta~\gamma\delta~\kappa\lambda~\rho\sigma}=0.
\end{equation}
Put another way, in three dimesions, the coefficients $C_{B_1\dots B_N}$ can depend on the second derivatives of $P^A$ only up to the third power. We may thus expand
\begin{eqnarray}
C_{B_1\dots B_N}&=&{}^{(3)}\Lambda_{B_1\dots B_N~Q\,\,\,\, R\,\,\,\, S}^{~~~~~~~~~~\alpha\beta~\gamma\delta~\kappa\lambda}\hat P^Q_{~~,\alpha\beta}~\hat P^R_{~~,\gamma\delta}~\hat P^S_{~~,\kappa\lambda}+{}^{(2)}\Lambda_{B_1\dots B_N~Q\,\,\,\, R}^{~~~~~~~~~~\alpha\beta~\gamma\delta}\hat P^Q_{~~,\alpha\beta}~\hat P^R_{~~,\gamma\delta}\\
&&+{}^{(1)}\Lambda_{B_1\dots B_N~Q }^{~~~~~~~~~~\alpha\beta}\hat P^Q_{~~,\alpha\beta}+{}^{(0)}\Lambda_{B_1\dots B_N},
\end{eqnarray}
where the coefficients ${}^{(i)}\Lambda$ can depend on the $P^A$ and their first derivatives only, and only the highest order coefficient ${}^{(3)}\Lambda$ must transform as a tensor density. In this way the dependence of the coefficients $C_{B_1\dots B_N}$ on the second derivatives of $\hat P^A$ can be completely eliminated from our differential equations. If $\Sigma$ is of higher dimension we simply have to add more derivatives in (\ref{Lambdas}). Thus, in general, the coefficients $C_{B_1\dots B_N}$ depend polynomially on the second derivatives of $\hat P^A$ at most to order $\dim \Sigma$. The coefficients now have to be determined from the remaining equations.
 
\subsection{Example: Canonical dynamics of second degree dispersion relations\label{sec_gr}}
We now illustrate how to solve the linear partial differential equations we identified in section \ref{sec_construction} in order to obtain gravitational dynamics, for the simplest case $\deg P = 2$. On a four-dimensional manifold, this directly yields Einstein-Hilbert gravitational dynamics with undetermined gravitational and cosmological constants (which appear as integration constants and must be fixed by experiment) as was first shown in \cite{HKT} a long time ago. The point here is of course that we have the relevant equations for any admissible dispersion relation, not only those of second degree, and only wish to illustrate that one can indeed proceed from these equations without further assumptions in order to obtain the gravitational dynamics of the specific spacetime geometry at hand. In particular, due to our foregoing comprehensive analysis that extracted all information from the constraint algebra, we do not need to draw on any results beyond our equations. 

In the case of a second rank tensor field $P$, which we consider here, all capital indices contain symmetric pairs of lower case greek indices running from $1$ to $3$. First we observe that the coefficients $M^{A\beta}$ and $Q_{B}{}^{A\beta}$ vanish since the non-local part of the Hamiltonian is equal to zero. Moreover, the coefficients $U^{A\alpha\beta}$ and $S^{A\beta}$ reduce to
\begin{equation}
U^{\alpha_1\alpha_2\beta\zeta} =- 2\hat P^{\beta(\alpha_1}\hat P^{\alpha_2)\zeta}\qquad\textrm{ and }\qquad S^{\alpha_1\alpha_2\beta} =- \hat P^{\beta\gamma}\partial_\gamma \hat P^{\alpha_1\alpha_2} +2 \hat P^{\gamma(\alpha_1}2\partial_\gamma \hat P^{\alpha_2)\beta}\,.
\end{equation}
Thus equation (\ref{maintwo}) for $N=2$ reads
\begin{equation}
0=C_{AB_1B_2}U^{A\beta\zeta}\,,
\end{equation}
which can be directly solved yielding $C_{AB_1B_2}=0$. Inserting this result into (\ref{maintwo}), starting with $N=4$ and iterating on all even $N$, we find that all coefficients $C_{B_1\dots B_N}$ with an odd number of capital indices greater or equal to three already vanish. For calculational convenience only, we perform a change of variables from $\hat P^{\alpha\beta}$ to $g_{\alpha\beta}$ with $\hat P^{\alpha\gamma}g_{\gamma\beta}=\delta^\alpha_\beta$ and substitute the first and second partial derivatives of $\hat P^{\alpha\beta}$ by those of $g_{\alpha\beta}$ accordingly. After this change of variables, equations (\ref{invarone}) and (\ref{symmi}) become
\begin{equation}
\frac{\partial C_{B_1\dots B_N}}{\partial g_{\alpha(\beta,\gamma\delta)}}=0\qquad\text{ and }\qquad\frac{\partial^2 C_{B_1\dots B_N}}{\partial g_{\alpha\beta,(\gamma\delta|}\partial g_{\mu\nu,\rho|\sigma)}}=0.
\end{equation}
Using a similar argument as in the previous section, we can now show that we even have
\begin{equation}
\frac{\partial^2C_{B_1\dots B_N}}{\partial g_{\alpha\beta,\gamma\delta}\partial g_{\mu\nu,\rho\sigma}}=0\,,
\end{equation}
because, in three dimensions, either one of the indices $1,2,3$ appears at least three times, so that all components of these tensor densities of weight one vanish according to the above symmetry conditions. Thus all remaining coefficients can depend at most linearly on the second derivatives of $g_{\alpha\beta}$. This has the direct consequence that according to equation (\ref{maintwo}) with $N=1$ the coefficient $C_{AB}$ cannot contain second derivatives of $g_{\alpha\beta}$. But then equation (\ref{maintwo}) implies $C_{B_1\dots B_N}=0$ for all even $N\geq 4$. Hence, it remains to determine $C_{AB}$, $C_A$ and $C$ to find the gravitational dynamics. We start with the discussion of the coefficient $C$. We already know that it has to be linear in the second derivatives of $g_{\alpha\beta}$ so that
\begin{equation}
C=C_0(g,\partial g)+\Lambda^{\alpha\beta\gamma\delta}_1g_{\alpha\beta,\gamma\delta}\,,
\end{equation}
where $\Lambda_1$ is a tensor density of weight one and  contains no second or higher partial derivatives of $g_{\alpha\beta}$. However, equation (\ref{invartwo}) implies that $\Lambda_1$, being a tensor density, can not even depend on the first partial derivatives of $g_{\alpha\beta}$. Thus equation (\ref{invarthreecontracted}) for $N=0$ can be rewritten into
\begin{equation}
C=\frac{2}{3}R_{\alpha\beta\gamma\delta}\Lambda^{\alpha\beta\gamma\delta}_1+\Lambda_0(g)\,,
\end{equation}
where $R_{\alpha\beta\gamma\delta}$ is the Riemann-Christoffel tensor of the metric $g_{\alpha\beta}$ and $\Lambda_0$ a tensor density of weight one that is solely constructed from $g_{\alpha\beta}$. In three dimensions, the Riemann tensor can of course be expressed in terms of the Ricci tensor $R_{\alpha\beta}=\hat P^{\gamma\delta} R_{\gamma\alpha\delta\beta}$. Now $\sqrt{-\det g}\, R$, where $R=R_{\alpha\beta}\hat P^{\alpha\beta}$ denotes the Ricci scalar, is the only weight-one scalar density linear in the second derivatives of $g_{\alpha\beta}$ that one may construct from the Ricci tensor and $g_{\alpha\beta}$, and the only scalar density of weight one one can construct from  $g_{\alpha\beta}$ alone is $\sqrt{-\det g}$. The minus sign under the square root accounts for the fact that with our normalisation condition the metric on the hypersurface is negative definite. Thus, we finally arrive at
\begin{equation}\label{potential}
C=-(2\kappa)^{-1} \sqrt{-\det g}\,(R-2\lambda)
\end{equation}
with two real integration constants $\kappa$ and $\lambda$. It is then simple to determine $C_{AB}$ from equation (\ref{maintwo}) for $N=1$ and we find
\begin{equation}\label{supermetric}
C_{\alpha\beta \mu\nu} =  (16\kappa)^{-1}\sqrt{-\det g} \left[
g_{\alpha \mu}g_{\beta \nu} + g_{\beta \mu}g_{\alpha \nu} - 2 g_{\alpha\beta}g_{\mu\nu}\right]\,.
\end{equation}
Finally, we calculate the coefficient $C_A$, which depends at most on the second derivatives of $g_{\alpha\beta}$ and is at most linear in those. Equation (\ref{mainfive}) reduces to
\begin{equation}
0=\hat P^{\alpha\beta}\hat P^{\gamma\delta}\nabla_{\alpha}C_{\beta\gamma},
\end{equation}
where we use the the torsion-free covariant derivative $\nabla_\alpha$ compatible with $g_{\alpha\beta}$ only for notational convenience. Using equations (\ref{invarone})-(\ref{invarthree}) and following a similar argument \cite{Lovelock} as for the coefficient $C$ immediately yields
\begin{equation}\label{oneindex}
C_{\alpha\beta}=\beta_1\sqrt{-\det g}\,(R_{\alpha\beta}-1/2g_{\alpha\beta} R)+\beta_2\sqrt{-\det g}\,g_{\alpha\beta}
\end{equation}
for some constants $\beta_1$ and $\beta_2$. The remaining equations (\ref{mainone}) with $N=2$ and (\ref{mainthree}) with $N=2$ involving $C_{\alpha\beta}$ are then identically satisfied. We note that the coefficient $C_{\alpha\beta}$ can be written as the functional gradient $\delta S /\delta \hat P^{\alpha\beta}$ of the scalar density $S=\beta_1\sqrt{-\det g}R-2\beta_2\sqrt{-\det g}$, and  finally make the transition from the full Lagrangian (\ref{Kseries}) to the superhamiltonian by means of the Legendre transformation (\ref{Legendreduals})-(\ref{reverse}). For the canonical momenta $\hat \pi_{\alpha\beta}$, one then has
\begin{equation}\label{transitionGR}
\hat \pi_{\alpha\beta}=\frac{\partial L}{\partial K^{\alpha\beta}}=2 C_{\alpha\beta\gamma\delta}K^{\gamma\delta}+\frac{\delta S}{\delta \hat P^{\alpha\beta}}.
\end{equation}
However, the canonical momenta (\ref{momentumvars}) are  only determined up an additive functional derivative of some scalar density of weight one with respect to $\hat P^{\alpha\beta}$. One can thus drop the second term on the right hand side of (\ref{transitionGR}) by redefining $\hat \pi_{\alpha\beta}\rightarrow\hat \pi_{\alpha\beta} - \delta S/\delta \hat P^{\alpha\beta}$ without changing the dynamics of the theory. Then the superhamiltonian reads
\begin{equation}\label{HamiltonGR}
H=C^{\alpha\beta\gamma\delta}\hat\pi_{\alpha\beta}\hat\pi_{\gamma\delta}-2C^{\alpha\beta\gamma\delta}C_{\alpha\beta}\hat\pi_{\gamma\delta}+(2\kappa)^{-1}\sqrt{-\det g}\,(R-2\lambda)
\end{equation}
with
\begin{equation}
C^{\alpha\beta\gamma\delta}=\frac{\kappa}{\sqrt{-\det g}}(\hat P^{\alpha\gamma}\hat P^{\beta\delta}+\hat P^{\beta\gamma}\hat P^{\alpha\delta}-\hat P^{\alpha\beta}\hat P^{\gamma\delta})\,.
\end{equation}
The second term in the superhamiltonian can be shown to be dynamically irrelevant \cite{Kuchar}, due to the special form of the coefficient $C_{\alpha\beta}$. With the superhamiltonian (\ref{HamiltonGR}) and the supermomentum $D_\alpha=2\hat P^{\beta\gamma}\nabla_{\beta}\hat \pi_{\alpha\gamma}$ from (\ref{supermom}), we have finally found (as \cite{HKT} did for a construction that only works for $\deg P=2$), the gravitational dynamics in the case of a three-dimensional hypersurface $\Sigma$ for a hyperbolic polynomial of degree two, also known as general relativity. The task to find canonical dynamics for dispersion relations beyond second degree is now of course to find solutions of our equations for $\deg P > 2$, which appears a much harder task. But this precisely what it takes if one wishes to consider modified dispersion relations in earnest. 

\newpage\section{Dynamics of tensorial spacetimes\label{sec_fieldgrav}}
We finally address the master problem of deriving the equations determining the gravitational dynamics of a fundamental geometric tensor field $G$, under the assumption that the latter gives rise to a hyperbolic, time-orientable and energy-distinguishing tensor field $P$ by virtue of specific matter field equations. This gravitational theory for $G$ is more fundamental than the phenomenological gravity dynamics derived for $P$ in the previous chapter, since fields couple directly to $G$, and so do point particles via $P$ constructed from $G$. But this greater generality comes at the price of a less sweeping construction scheme. While the always totally symmetric, even rank tensor fields $P$ can be treated in precisely the same way for any rank, the fundamental geometric tensors $G$ come in all possible ranks and symmetries (as long as one can couple matter fields to them), and the construction of their geometric phase space must proceed in fashion of a case-by-case analysis. 
But apart from these technical details, the overall construction is as simple and inevitable as in the previous chapter, and one obtains also a system of homogeneous linear partial differential equations whose solutions determine the gravitational dynamics of the geometry $(M,G)$.

\subsection{Construction of tensorial spacetime geometries and their dynamics}
The construction of gravitational dynamics for a fundamental geometric tensor field $G$ proceeds logically exactly along the same lines as that for dispersion relations. The only relevant difference consists in the choice of canonical variables for the dynamics, and all the technical modifications this entails. To separate the essential steps from their technical details, we therefore quickly prescribe the general recipe one has to follow to make any candidate geometry $(M,G)$ into a spacetime structure and to derive the equations determining their canonical dynamics. We will then see this recipe in action in the next two sections.  
\begin{enumerate}
\item Decide on a tensor field $G$ of arbitrary valence as the geometry on a smooth manifold $M$.\\[6pt]
More generally, one may also choose a collection of tensor fields $G=(G_1, G_2, \dots)$, each possibly of different valence, to provide the geometry. This would be the case, for instance, if one aimed at studying a bosonic string background featuring a metric $g$, a two-form field $B$ and a scalar $\phi$, say, using the philosophy of this paper.
\item Decide on matter dynamics to define the causal structure impressed on the geometry.\\[6pt]
These matter equations may well be of phenomenological nature, as were the Maxwell equations before 1905. This is where the theory gets its vital injection from realistic physics. 
\item Calculate the totally symmetric covariant tensor field $P_G$ associated with the linear(ized) matter field equations in terms of the geometry $G$.\\[6pt]
This is straightforward in principle, but may in practice require to first remove gauge ambiguities. If several matter field equations are present in the theory, one needs to consider their entirety to calculate $P$.
\item Restrict attention to geometries $(M,G)$ for which $P_G$ is hyperbolic, time-orientable and energy-distinguishing.\\[6pt]
Only these deserve to be called spacetimes. In order to get an overview over which algebraic classes of the geometry $G$ present spacetimes, it is often useful to figure out the algebraic classification of the geometric tensor $G$ under $\textrm{GL}(\dim M)$ transformations and associated normal forms. 
\item Construct the configuration variables describing the spatial geometry on an accessible initial data hypersurface by normal and tangential projections, eliminate the degrees of freedom fixed by $P_G=1$ and $P_G^\alpha=0$, and associate canonically conjugate momenta to all remaining degrees of freedom.
\item Construct the supermomentum and superhamiltonian exactly along the same lines as done in section \ref{sec_ppgrav}, but with the spatial point particle geometry replaced by that for fields, as we will illustrate for area metric spacetimes in four dimensions in the following two sections.\\[6pt]
The precise form of the coefficients will depend heavily on the geometry $G$ chosen. But once a concrete geometric tensor $G$ is chosen, and its independent degrees of freedom have been identified, the calculation goes through also in this case without complications.
\item Solve the resulting system of linear partial differential equations to determine the local part of the superhamiltonian.\\[6pt]
How difficult this is now very much varies with the geometry $G$ that has been chosen.
\end{enumerate}
For the simple case of metric geometry carrying Maxwell theory, execution of this programme leads to the condition that the metric must have Lorentzian signature, and the  system of homogeneous linear partial differential equations has a unique family of solutions, giving rise to the standard Einstein-Hilbert gravitational action with undetermined gravitational and cosmological constants (which appear as constants of integration). Essentially, this has been shown a long time ago \cite{HKT}, and is of course recovered as a very special case of our general construction. 

Any other tensorial geometry requires a separate case-by-case analysis for virtually all of the above steps. We therefore choose to illustrate the procedure for area metric geometry, which accompanied us throughout this paper as a particularly interesting example for the workings of our general theory. 
   
\subsection{Phase space for area metric geometry seen by electromagnetic fields}\label{sec_fieldphasespace}
To illustrate the procedure of finding canonical dynamics directly for a fundamental geometric tensor field $G$ underlying a chosen field theory, we will concentrate, for definiteness, on the particular example of a four-dimensional area metric geometry coupled to electromagnetric fields. We assume that the inverse area metric $G^{abcd}$ is everywhere non-cyclic such that with (\ref{PArea}) the totally symmetric geometric tensor $P_G^{abcd}$ takes the form
\begin{equation}
\label{normArea}
P_G^{abcd}=-\frac{24}{(G^{ijkl}\epsilon_{ijkl})^2}\epsilon_{mnpq}\epsilon_{rstu}G^{mnr(a}G^{b|ps|c}G^{d)qtu}.
\end{equation}
Using the complete spacetime covector frame $\{n,\epsilon^1,\epsilon^2,\epsilon^3\}$ constructed from $P$ along a hypersurfaces $X(\Sigma)$ given in terms of the embedding map $X(y)$, we then define the functionals
\begin{eqnarray}
G^{\alpha \beta}(y)[X] &=& G(n(y),\epsilon^{\alpha}(y), n(y), \epsilon^{\beta})\,,\\
G^{\alpha}_{~~\beta}(y)[X] &=&\frac{1}{2}\omega_{G\beta\gamma\delta}\, G(n(y),\epsilon^{\alpha}(y), \epsilon^{\gamma}(y), \epsilon^{\delta}(y))\,,\\
G_{\alpha \beta}(y)[X] &=&\frac{1}{4}\omega_{G\alpha\gamma\delta}\,\omega_{G\beta\mu\nu} G(\epsilon^{\gamma}(y), \epsilon^{\delta}(y), \epsilon^{\mu}(y), \epsilon^{\nu}(y))\,.
\end{eqnarray}
where we used the volume form $\omega_{G\alpha\beta\gamma}=(-\det G^{\alpha\beta})^{-1/2}\epsilon_{\alpha\beta\gamma}$ induced by the symmetric hypersurface tensor field $G^{\alpha\beta}$ to construct the hypersurface tensor fields $G^{\alpha}_{~~\beta}$ and $G_{\alpha\beta}$ from the other possible projections of the inverse area metric onto the hypersurface $X(\Sigma)$. Note that the index positions really distinguish unrelated tensor fields, which together encode the degrees of freedom of the inverse area metric on the hypersurface. From the normalisation conditions $P_G(n,n,n,n)=1$ and $P_G(n,n,n,\epsilon^{\alpha})=0$ it follows that $G^{\alpha}_{~~\beta}$ can be assumed to be trace-free and symmetric with respect to $G^{\alpha\beta}$. 

The phase space of a four-dimensional area metric spacetime is then spanned by tensor fields
\begin{equation}\label{areaVar}
\hat G^{\alpha\beta},\qquad \hat G^{\alpha}_{~~\beta},\qquad \hat G_{\alpha\beta}
\end{equation}
on the three-dimensional manifold $\Sigma$ and their respective canonical momenta
\begin{equation}\label{AreaPi}
\hat \Pi_{\alpha\beta}, \qquad \hat \Pi_{\alpha}^{~~\beta},\qquad \hat \Pi^{\alpha\beta},
\end{equation}
which are taken to be tensor densities of weight one.
Adjoining canonical momenta again is then equivalent to adopting a Poisson bracket
\begin{equation}
\{\hat E,\hat F\}=\int dy \left[\frac{\delta \hat E}{\delta \hat G^A}\frac{\delta \hat F}{\delta \Pi_A}-\frac{\delta \hat F}{\delta \hat G^A}\frac{\delta \hat E}{\delta \Pi_A}\right]
\end{equation}
on the space of functionals of the canonical variables $\hat G^A$ and $\hat \Pi_A$, where the capital index $A$ collectively denotes the different greek indices with their respective positions: ${}^A=({}^{\alpha\beta},{}^{\alpha}_{~~\beta},{}_{\alpha\beta})$.

Again we will look for dynamics in terms of a Hamiltonian as it appears in (\ref{classHam}) that evolves the phase space variables $(\hat G^A,\hat \Pi_A)$ with an evolution parameter $t$, such that the embedding of the data at time $t$ by virtue of a foliation $X_t:\Sigma\rightarrow M$ produce an inverse area metric $G^{abcd}$ on $M$ whose dispersion relation is hyperbolic, time-orientable and energy distinguishing.
The 21 components of the inverse area metric can then be reconstructed from the 17 independent components of the symmetric tensor field $\hat G^{\alpha\beta}$, the trace-free hypersurface tensor field $\hat G^{\alpha}_{~~\beta}$ which is symmetric with respect to $\hat G^{\alpha\beta}$, the symmetric hypersurface tensor field $\hat G_{\alpha\beta}$ and the spacetime vector frame $\{T_t,e_{t\,\alpha}\}$ by
\begin{eqnarray}
G^{abcd}[X_t(y)]&=&4\hat G_t^{\beta\delta} T_t^{[a}e^{b]}_{t\,\beta}T_t^{[c}e^{d]}_{t\,\delta} +\hat G_{t\,\rho\sigma}(\omega_G^{-1})^{\rho\alpha\beta}(\omega_G^{-1})^{\sigma\gamma\delta}e^{a}_{t\,\alpha} e^{b}_{t\,\beta} e^{c}_{t\,\gamma} e^{d}_{t\,\delta}\nonumber\\
&&+2(\hat G^{\beta}_{t~\rho}+\delta^\beta_{~~\rho})(\omega_G^{-1})^{\rho\gamma\delta}T_t^{[a}e^{b]}_{t\,\beta}e^{c}_{t\,\gamma} e^{d}_{t\,\delta}.
\end{eqnarray}
The conceptual steps in the construction of the supermomentum $\hat{\mathcal{D}}$ and the superhamiltonian $\hat{\mathcal{H}}$ on the phase space given by (\ref{areaVar}) and (\ref{AreaPi}) are precisely the same as for the pure point particle geometry in section \ref{sec_ppgrav}. We will quickly go through these steps in the next section.

\subsection{Canonical dynamics for area metric spacetime}\label{sec_candyn}
We already saw in the previous section that the canonical phase space in the case of a four-dimensional area metric spacetime consists of the tensor fields $\hat G^{\alpha\beta}$ (symmetric), $\hat G^{\alpha}_{~~\beta}$ (trace-free, symmetric with respect to $\hat G^{\alpha\beta}$) and $\hat G_{\alpha\beta}$ (symmetric) as well as their conjugate momenta $\hat \Pi_{\alpha\beta}$, $\hat \Pi_{\alpha}^{~~\beta}$ and $\hat \Pi^{\alpha\beta}$ with the same respective algebraic properties. The superhamiltonian $\hat{\mathcal{H}}(N)$ and the supermomentum $\hat{\mathcal{D}}(N^{\alpha}\partial_{\alpha})$ satisfy the Poisson algebra relations (\ref{repHH}), (\ref{repDH}) and (\ref{repDD}), but now the symbol $\hat P_G^{\alpha\beta}$ on the right hand side of (\ref{repHH}) is not a canonical variable itself, but the particular phase space function
\begin{equation}\label{AreaMetricP}
\hat P_G^{\alpha\beta}=\frac{1}{6}\left(\hat G^{\alpha\beta} \hat G^{\gamma\delta}\hat G_{\gamma\delta}-\hat G^{\alpha\gamma}\hat G^{\delta\beta} \hat G_{\gamma\delta}-2 \hat G^{\alpha\beta}\hat G^{\gamma}_{~~\delta}\hat G^{\delta}_{~~\gamma}+3\hat G^{\gamma\delta}\hat G^{\alpha}_{~~\gamma}\hat G^{\beta}_{~~\delta}\right).
\end{equation}
The construction of the supermomentum follows the same steps as in the case of the point particle geometry. One readily finds
\begin{eqnarray}
\hat{\mathcal{D}}(N^{\gamma}\partial_\gamma)=\int dy N^{\gamma}(y)&\Big[&(\partial_{\gamma}\hat G^{\alpha\beta})\hat \Pi_{\alpha\beta}+2\partial_\alpha(\hat G^{\alpha\beta}\hat \Pi_{\beta\gamma})+(\partial_{\gamma}\hat G_{\alpha\beta})\hat \Pi^{\alpha\beta}-2\partial_\alpha(\hat \Pi^{\alpha\beta}\hat G_{\beta\gamma})\\
&&+(\partial_\gamma\hat G^{\alpha}_{~~\beta})\hat \Pi^{~~\beta}_{\alpha}+\partial_\alpha(\hat G^{\alpha}_{~~\beta}\hat\Pi^{~~\beta}_{\gamma})-\partial_{\alpha}(\hat G^{\beta}_{~~\gamma}\hat\Pi^{~~\alpha}_{\beta}) \Big].
\end{eqnarray}
The non-local part $\hat{\mathcal{H}}_{\text{non-local}}$ of the superhamiltonian can be found using the first of the necessary relations (\ref{application}). One calculates
\begin{eqnarray}
\frac{\hat{\mathcal{H}}(N)}{\hat \Pi_{\alpha\beta}(z)}&=&N(z)[\dots]+\partial_\gamma N(z)\left[-2(\omega_{\hat G}^{-1})^{\delta\gamma(\alpha}\hat G^{\beta)}_{~~\delta}\right]\,,\\
\frac{\hat{\mathcal{H}}(N)}{\hat \Pi^{~~\beta}_{\alpha}(z)}&=&N(z)[\dots]+\partial_\gamma N(z)\left[-3\,\omega_{\hat G\beta\sigma\tau} \hat G^{\alpha\sigma}\hat P_G^{\tau\gamma}-(\omega_{\hat G}^{-1})^{\gamma\alpha\sigma}\hat G_{\sigma\beta}\right]\,,\\
\frac{\hat{\mathcal{H}}(N)}{\hat \Pi^{\alpha\beta}(z)}&=&N(z)[\dots]+\partial_\gamma N(z)\left[-6\,\omega_{\hat G\sigma\tau(\beta}\hat G^{\sigma}_{~~\alpha)}\hat P_G^{\tau\gamma}\right]\,,
\end{eqnarray}
which may be integrated to yield the non-local part of the superhamiltonian
\begin{eqnarray}
\hat{\mathcal{H}}_{\text{non-local}}&=&2\partial_{\gamma}\left[(\omega_{\hat G}^{-1})^{\delta\gamma(\alpha}\hat G^{\beta)}_{~~\delta}\hat\Pi_{\alpha\beta}\right](y)+3\partial_{\gamma}\left[\omega_{\hat G\beta\sigma\tau} \hat G^{\alpha\sigma}\hat P_G^{\tau\gamma}\hat\Pi^{\beta}_{~~\alpha}\right](y)\\
&+&\partial_{\gamma}\left[(\omega_{\hat G}^{-1})^{\gamma\alpha\sigma}\hat G_{\sigma\beta}\hat \Pi^{\beta}_{~~\alpha}\right](y)+6\partial_{\gamma}\left[\omega_{\hat G\sigma\tau(\beta}\hat G^{\sigma}_{~~\alpha)}\hat P_G^{\tau\gamma}\hat\Pi^{\alpha\beta}\right](y).
\end{eqnarray}
It remains to evaluate the Poisson bracket of two superhamiltonians to determine its local part. The analysis proceeds along precisely the same lines as in the case of the pure point particle geometry. We perform a Legendre transformation of the local part $\hat{\mathcal{H}}_{local}$ of the superhamiltonian with respect to the momenta $\hat\Pi_A$. The resulting Lagrangian $L[\hat G^A](K^A)$ then satisfies the linear functional differential equation
\begin{eqnarray}\label{homolinArea}
   0 &=& - \frac{\delta L(x)}{\delta \hat{G}^A(y)} K^A(y) + \partial_{y^\zeta}\left[\frac{\delta L(x)}{\delta \hat{G}^A(y)} M^{A\zeta}(y)\right] - \frac{\partial L(x)}{\partial K^A(x)}K^B(x)Q_{B}{}^{A\beta}(x)\partial_\beta\delta_x(y)\nonumber\\
&&+\frac{\partial L(x)}{\partial K^A(x)}\left[ U^{A\mu\nu}(x)\partial^2_{\mu\nu} \delta_x(y) + S^{A\mu}(x) \partial_\mu \delta_x(y) \right] - (x \leftrightarrow y). 
\end{eqnarray}
The coefficients $M^{A\beta}$ and $Q^{~~A\beta}_{B}$ can be read off from the functional derivatives of the non-local part of the superhamiltonian with respect to the canonical variables. The coefficients $U^{A\gamma\delta}$ read
\begin{eqnarray}
U^{\alpha\beta\gamma\delta}&=&-6\hat P^{(\gamma|(\alpha}\hat G^{\beta)|\delta)},\\
U^{\alpha}_{~~\beta}{}^{\gamma\delta}&=&-3\hat P^{\alpha(\gamma}\hat G^{\delta)}_{~~\beta}+3 \hat P^{\sigma(\gamma}\delta^{\delta)}_{\beta}\hat G^{\alpha}_{~~\sigma},\\
U_{\alpha\beta}^{~~~\gamma\delta}&=&6\,\hat G_{\sigma(\alpha}\delta^{(\gamma}_{\beta)}\hat P^{\delta)\sigma}.
\end{eqnarray}
The coefficients $S^{A\gamma}$ can be calculated from
\begin{equation}
S^{A\gamma}=\partial_\beta Q_B^{~~A(\beta|}M^{B|\gamma)}-Q_{B}^{~~A[\beta|}\partial_\beta M^{B|\gamma]}-\partial_\beta U^{A\beta\gamma}-3 \hat P^{\gamma\beta}\partial_\beta \hat G^A-3V^{A\gamma},
\end{equation}
with $V^{\alpha\beta\gamma}=2\hat P^{\gamma(\alpha}\partial_{\delta}\hat G^{\beta)\delta}$, $V^{\alpha~~\gamma}_{~~\beta}=\hat P^{\gamma\alpha}\partial_\delta \hat G^{\delta}_{~~\beta}-\hat P^{\gamma\delta}\partial_\beta \hat G^{\alpha}_{~~\delta}$ and $V_{\alpha\beta}^{~~~\gamma}=-2P^{\gamma\delta}\partial_{(\alpha}\hat G_{\beta)\delta}$. Expanding the Lagrangian $L[\hat G^A](K^A)$ into a power series in the velocities $K^A$,
\begin{equation}
L(x)[\hat G^A](K^A)=\sum_{i=0}^{\infty}C(x)[\hat G^A]_{B_1\dots B_i}K^{B_1}\dots K^{B_i},
\end{equation}
one derives exactly the same equations (\ref{mainone})-(\ref{mainfive}) for the coefficients $C_{B_1\dots B_N}$ as in the point particle case. The coefficients $C_{B_1\dots B_N}$ are again tensor densities of weight one and, as a result of the algebra equations, depend at most on the second partial derivatives of the fields $\hat G^A$. Since the hypersurface $\Sigma$ is of dimension three, it again follows that the coefficients depend on the second partial derivatives only up to the third power.

The invariance equations following from the transformation properties of the weight-one densities $C_{B_1\dots B_N}$ (or, fully equivalently, from the Poisson bracket of the supermomentum with the superhamiltonian) can be derived in straightforward fashion. The first invariance identity takes the form
\begin{equation}
0=2\hat G^{\mu(\alpha}\frac{\partial C_{B_1\dots B_N}}{\partial \partial^2_{\beta\gamma)}\hat G^{\mu\rho}}+\hat G^{(\alpha}_{~~\mu}\frac{\partial C_{B_1\dots B_N}}{\partial \partial^2_{\beta\gamma)}\hat G^{\rho}_{~~\mu}}-\hat G^{\mu}_{~~\rho}\frac{\partial C_{B_1\dots B_N}}{\partial \partial^2_{(\alpha\beta}\hat G^{\mu}_{~~\gamma)}}-2\hat G_{\rho\mu}\frac{\partial C_{B_1\dots B_N}}{\partial \partial^2_{(\alpha\beta}\hat G_{\gamma)\mu}}.
\end{equation}
The second invariance equation reads
\begin{eqnarray}
0&=&2\hat G^{\mu(\alpha}\frac{\partial C_{B_1\dots B_N}}{\partial \partial_{\beta)}\hat G^{\mu\rho}}+4\partial_\nu\hat G^{\mu(\alpha}\frac{\partial C_{B_1\dots B_N}}{\partial \partial^2_{\beta)\nu}\hat G^{\mu\rho}}-\partial_\rho\hat G^{\mu\nu}\frac{\partial C_{B_1\dots B_N}}{\partial \partial^2_{\alpha\beta}\hat G^{\mu\nu}}\nonumber\\
 &+&\hat G^{(\alpha}_{~~\mu}\frac{\partial C_{B_1\dots B_N}}{\partial \partial_{\beta)}\hat G^{\rho}_{~~\mu}}-\hat G^{\mu}_{~~\rho}\frac{\partial C_{B_1\dots B_N}}{\partial \partial_{(\alpha}\hat G^{\mu}_{~~\beta)}}+2\partial_\nu\hat G^{(\alpha}_{~~\mu}\frac{\partial C_{B_1\dots B_N}}{\partial \partial^2_{\beta)\nu}\hat G^{\rho}_{~~\mu}}-2\partial_\nu\hat G^{\mu}_{~~\rho}\frac{\partial C_{B_1\dots B_N}}{\partial \partial^2_{\nu(\alpha}\hat G^{\mu}_{~~\beta)}}\nonumber\\
 &-&\partial_\nu\hat G^{\mu}_{~~\rho}\frac{\partial C_{B_1\dots B_N}}{\partial \partial^2_{\alpha\beta}\hat G^{\mu}_{~~\nu}}-2\hat G_{\rho\mu}\frac{\partial C_{B_1\dots B_N}}{\partial \partial_{(\alpha}\hat G_{\beta)\mu}}-4\partial_\nu\hat G_{\mu\rho}\frac{\partial C_{B_1\dots B_N}}{\partial \partial^2_{\nu(\alpha}\hat G_{\beta)\mu}}-\partial_\rho \hat G_{\mu\nu}\frac{\partial C_{B_1\dots B_N}}{\partial \partial^2_{\alpha\beta}\hat G_{\mu\nu}}.
\end{eqnarray}
The last invariance identity is even more complicated and we only display its contracted form:
\begin{eqnarray}
-(3+n)C_{B_1\dots B_N}&=&2\hat G^{\mu\nu}\frac{\partial C_{B_1\dots B_N}}{\partial \hat G^{\mu\nu}}-2\hat G_{\mu\nu}\frac{\partial C_{B_1\dots B_N}}{\partial \hat G_{\mu\nu}}\nonumber\\
&+&\partial_\rho\hat G^{\mu\nu}\frac{\partial C_{B_1\dots B_N}}{\partial\partial_\rho \hat G^{\mu\nu}}-\partial_\rho\hat G^{\mu}_{~~\nu}\frac{\partial C_{B_1\dots B_N}}{\partial\partial_\rho \hat G^{\mu}_{~~\nu}}-3\partial_\rho \hat G_{\mu\nu}\frac{\partial C_{B_1\dots B_N}}{\partial\partial_\rho \hat G_{\mu\nu}}\nonumber\\
&-&2\partial^2_{\rho\sigma}\hat G^{\mu}_{~~\nu}\frac{\partial C_{B_1\dots B_N}}{\partial\partial^2_{\rho\sigma} \hat G^{\mu}_{~~\nu}}-4\partial^2_{\rho\sigma}\hat G^{\mu\nu}\frac{\partial C_{B_1\dots B_N}}{\partial\partial^2_{\rho\sigma} \hat G_{\mu\nu}},
\end{eqnarray}
where $n$ is the difference of the total number of subscript indices and the total number of superscript indices in the coefficients $C_{B_1\dots B_N}$.

\newpage\section{Conclusions\label{sec_conclusions}}\enlargethispage{32pt}
In this paper, we addressed the question of how to construct canonical gravitational dynamics for spacetime geometries beyond the Lorentzian manifolds featuring in Einstein's general relativity. 

The first step consisted in an analysis of what kind of tensor fields $G$ on a smooth manifold $M$ can serve as a spacetime geometry in the first place, dependent on the presence of specific matter field dynamics. Indeed, the geometry must be such that all matter field equations are predictive, interpretable and quantizable. These conditions on the matter field dynamics impose three corresponding algebraic conditions on a totally symmetric tensor field $P$, which is defined in terms of the fundamental geometric tensor field $G$ and whose precise form arises from the matter field dynamics: $P$ needs to be hyperbolic, time-orientable and energy-distinguishing, as reviewed in section \ref{sec_primer}. So in order to start the whole machinery presented here, we first need to know which matter couples in which way to the tensorial geometry. We do not see this as a weakness of the formalism, but rather as an insight; it was Maxwell theory that justified Einstein to promote Lorentzian manifolds to the status of a  spacetime geometry, \new{and experimental observation of any matter that does not mimick the structure of Maxwell theory (non-half-integer spin or superluminal matter, for instance)} will force us to choose another tensorial geometry. But certainly one that is hyperbolic, time-orientable and energy-distinguishing. Fortunately, with the results of chapter \ref{sec_kinematics}, we have all these geometries under excellent technical control. 

Directly from these purely kinematical insights, one can calculate the deformation algebra of hypersurfaces in any hyperbolic, time-orientable and energy-distinguishing geometry. This is the algebra of linear operators that describe how the geometry induced on a  hypersurface changes when the hypersurface is deformed in normal and tangential directions. And gravitational dynamics is precisely about understanding these changes in the geometry on initial data surfaces, as has been clarified in seminal work of Hojmann, Kuchar and Teitelboim for the special case of Lorentzian manifolds, building on the canoncial formalism introduced by Arnowitt, Deser and Misner. The most important result of chapter \ref{sec_kinematics}, from a practical point of view, is therefore that one can calculate the deformation algebra of hypersurfaces in any hyperbolic, time-orientable and energy-distinguishing tensorial geometry.
This is by no means trivial, since the existence and uniqueness of the way to associate normal directions along a hypersurface with its canonical normal co-directions by means of a Legendre map requires all three algebraic properties: the hyperbolicity, time-orientability and energy-distinguishing property. Despite appearances, this also applies to metric geometry (where any one of these conditions is equivalent to the requirement of a Lorentzian signature), since although, purely formally, one can still construct normal directions from normal co-directions for other signatures, they lose their physical meaning.

Canonical gravitational dynamics for the spatial geometry that are ultimately invariant under spacetime diffeomorphisms must be given by a pure constraint Hamiltonian (which is of course a functional of the geometric degrees of freedom and associated conjugate momenta on an initial data hypersurface) composed of two separate first class constraints---corresponding to spatial diffeomorphism invariance within the hypersurface on the one hand, and invariance under diffeomorphisms away from the hypersurface on the other hand. The r\^ole of the deformation algebra, in this geo\-metrodynamic language, is that these constraints must satisfy canonical Poisson bracket relations of the same form as the commutator algebra of the normal and tangential deformation operators on functionals of the hypersurface embedding map. The task is thus to determine the constraint functionals from this Poisson algebra.

This would be a mere representation theory problem if the Poisson algebra were a Lie algebra. But only two of the three bracket relations feature structure constants, and their impact on the form of the constraint functionals is thus readily established. The third Poisson bracket relation, however, features a structure function that captures the impact of the particular hyperbolic, time-orientable and energy-distinguishing tensorial geometry to which one wishes to give dynamics. Determining the implications of this bracket amounts, at first sight, to the truly daunting task of solving a system of non-linear functional-differential equations. The better part of chapters \ref{sec_ppgrav} and \ref{sec_fieldgrav} is thus devoted to reducing this to the equivalent, and principally manageable, problem of solving a system of homogeneous linear partial differential equations. And this set of equations already contains, by construction, all possible classical gravitational dynamics one can give to a tensorial spacetime geometry that can carry predictive, interpretable and quantizable matter fields. In the philosophy of this paper, the physical problem of finding diffeomorphism-invariant gravity theories alternative to Einstein's general relativity is shown to be equivalent to the mere mathematical task of solving these linear partial differential equations.

This casts important physical questions into precise mathematical form. The question whether there are any alternatives to general relativity turns into the problem of existence of solutions; the question whether there is a choice between various dynamics for a given tensorial spacetime geometry translates into the question of their uniqueness; and finally, the actual construction of all concrete gravitational dynamics amounts to nothing more, but also nothing less, than explicitly finding the exact solutions of these linear partial differential equations.

The difference between the treatments in chapter \ref{sec_ppgrav} and \ref{sec_fieldgrav} is that only in the latter do we construct dynamics for the  fundamental tensorial spacetime geometry $G$ to which also fields can couple, while in the former we give dynamics only to the totally symmetric tensor field $P$ seen by point particles. 
In the special case of Lorentzian geometry, the two points of view accidentally coincide, since the tensor field $P$ encodes precisely the same degrees of freedom as the fundamental Lorentzian metric $g$ to which fields can couple. 
The key result of chapter \ref{sec_ppgrav} is that there is at most a one-integer-family of essentially different gravity theories that differ in their prediction of particle trajectories, and we wrote down the complete set of equations for all these theories. So if one is interested in the motion of massive and massless point matter only, one can ignore which particular fundamental geometric structure underlies a hyperbolic, time-orientable and energy-distinguishing dispersion relation, and compare observational data with these phenomenological theories. In contrast, if one wishes to consider a full gravitational theory to which both point particles and fields can couple, one needs to construct these along the lines laid out in chapter \ref{sec_fieldgrav}. The resulting theories are more fundamental, but this comes at the price that the equations yielding their dynamics depend more heavily on the specific type of tensor field $G$ and require the explicit specification of the predictive, interpretable and quantizable matter coupling to it. Our derivation of the relevant equations for four-dimensional area metric geometry carrying general linear electrodynamics at the end of chapter \ref{sec_fieldgrav}, however, shows that also this more fundamental programme can be executed.

The main open question is how to find solutions to the system of homogeneous linear partial differential equations in either case. But this will well be worth the effort, since solving these equations immediately allows to answer a string of pertinent physical questions in gravity theory. Four questions of high relevance for a number of current research programmes are how to 
\begin{itemize}
\item[(i)] settle the issue of which modified dispersion relations are admissible, and how they are determined dynamically, in order to conduct a focused search for experimental signatures. 
\item[(ii)] provide canonical dynamics to one's favourite candidate of a tensorial geometry without further assumptions, starting from matter dynamics coupling to this geometry \footnote{This certainly presents an entirely new angle on---and more importantly: physically well-founded approach to---the construction of gravitational dynamics for a non-symmetric metric or for string geometries determined, for instance, by a metric $g$, a two-form potential $B$ and dilaton $\phi$ from a purely canonical point of view.}. 
\item[(iii)]  free the evaluation of observational raw data from the confines of Lorentzian geometry and Einstein dynamics, in favour of the wider framework that includes all hyperbolic, time-orientable and energy-distinguishing tensorial geometries, which one is led to consider in the light of matter dynamics that would qualify as non-causal in Lorentzian spacetime.   
\item[(iv)] reveal all possible classical limits of quantum gravity theories where the fundamental geometric structure can be expressed in terms of tensor fields \footnote{For classical theories based on a connection formulation, this may require a reformulation in terms of purely tensorial objects whenever possible. Possibly some pure connection theories therefore escape the formalism presented here, but in this case we feel a careful evaluation of their kinematical apparatus, in particular the definition of observer frames, should yield the relevant information. This then certainly requires a case-by-case analysis.}.  
\end{itemize}
Our future progress on these questions thus hinges on solving the equations derived in this work.

\begin{acknowledgments}
The authors gratefully acknowledge instructive discussions with, and most insightful comments by, Thomas Thiemann, Domenico Giulini, Claus L\"ammerzahl, Volker Perlick, Klaus Mecke, Kirill Krasnov, Sergio Rivera and Dennis R\"atzel.    
KG and FPS thank the Nordic Institute for Theoretical Physics in Stockholm and the Excellence Cluster Universe in Munich, and KG additionally the Albert Einstein Institute, for their hospitality and support where parts of this work have been completed. 
MNRW gratefully acknowlegdes support through the
Emmy Noether Fellowship grant WO 1447/1-1. CW thanks the Studienstiftung des deutschen Volkes and the International Max Planck Research School for Geometric Analysis, Gravitation and String Theory for their support.
\end{acknowledgments}

\end{document}